\documentclass[twocolumn,aps,prc,superscriptaddress,amssymb]{revtex4}
\usepackage{graphicx}
\usepackage[english]{babel}
\usepackage{amssymb}
\usepackage[latin2]{inputenc}
\usepackage[T1]{fontenc}
\usepackage{latexsym}
\usepackage{here}
\usepackage{mwe}
\usepackage{subscript}
\usepackage{longtable}
\usepackage{textcomp}
\usepackage{amsmath}
\usepackage{amssymb}
\usepackage{graphicx}
\usepackage{epstopdf}
\usepackage{epsfig}
\usepackage{dcolumn}
\usepackage{bm}
\usepackage[modulo]{lineno}
\usepackage{amssymb}
\usepackage{subfigure}
\usepackage{multirow}
\usepackage[colorlinks=true,urlcolor=blue]{hyperref}
\begin{document}
\title
{First $\boldsymbol{\beta}$-decay spectroscopy of \textsuperscript{135}In and new $\boldsymbol{\beta}$-decay branches of \textsuperscript{134}In}
%
%
\author{M.\,Piersa-Si\l{}kowska}
\affiliation{Faculty of Physics, University of Warsaw, PL 02-093 Warsaw, Poland}
\author{A.\,Korgul}
\affiliation{Faculty of Physics, University of Warsaw, PL 02-093 Warsaw, Poland}
\author{J.\,Benito}
\affiliation{Grupo de F\'isica Nuclear and IPARCOS, Universidad Complutense de Madrid, CEI Moncloa, E-28040 Madrid, Spain}
\author{L.\,M.\,Fraile}
\affiliation{Grupo de F\'isica Nuclear and IPARCOS, Universidad Complutense de Madrid, CEI Moncloa, E-28040 Madrid, Spain}
\affiliation{CERN, CH-1211 Geneva 23, Switzerland}
\author{E.\,Adamska}
\affiliation{Faculty of Physics, University of Warsaw, PL 02-093 Warsaw, Poland}
\author{A.\,N.\,Andreyev}
\affiliation{Department of Physics, University of York, York, YO10 5DD, United Kingdom}
%
\author{R.\,\'Alvarez-Rodr\'iguez}
\affiliation{Escuela Tecnica Superior de Arquitectura, Universidad Polit\'ecnica de Madrid, E-28040 Madrid, Spain}
\author{A.\,E.\,Barzakh}
\affiliation{Petersburg Nuclear Physics Institute, NRC Kurchatov Institute, 188300 Gatchina, Russia}
\author{G.\,Benzoni}
\affiliation{Istituto Nazionale di Fisica Nucleare, Sezione di Milano, I-20133 Milano, Italy}
\author{T.\,Berry}
\affiliation{Department of Physics, University of Surrey, Guildford GU2 7XH, United Kingdom}
\author{M.\,J.\,G.\,Borge}
\affiliation{CERN, CH-1211 Geneva 23, Switzerland}
\affiliation{Instituto de Estructura de la Materia, CSIC, E-28006 Madrid, Spain}
\author{M.\,Carmona}
\affiliation{Grupo de F\'isica Nuclear and IPARCOS, Universidad Complutense de Madrid, CEI Moncloa, E-28040 Madrid, Spain}
\author{K.\,Chrysalidis}
\affiliation{CERN, CH-1211 Geneva 23, Switzerland}
\author{J.\,G.\,Correia}
\affiliation{CERN, CH-1211 Geneva 23, Switzerland}
\affiliation{C2TN, Centro de Ci\^encias e Tecnologias Nucleares, Instituto Superior T\'ecnico, Universidade de Lisboa, Portugal}
\author{C.\,Costache}
\affiliation{``Horia Hulubei'' National Institute of Physics and Nuclear Engineering, RO-077125 Bucharest, Romania}
\author{J.\,G.\,Cubiss}
\affiliation{CERN, CH-1211 Geneva 23, Switzerland}
\affiliation{Department of Physics, University of York, York, YO10 5DD, United Kingdom}
\author{T.\,Day\,Goodacre}
\affiliation{CERN, CH-1211 Geneva 23, Switzerland}
\affiliation{School of Physics Astronomy, The University of Manchester, Manchester M13 9PL, United Kingdom}
\author{H.\,De Witte}
\affiliation{KU Leuven, Instituut voor Kern- en Stralingsfysica, Celestijnenlaan 200D, 3001 Leuven, Belgium}
\author{D.\,V.\,Fedorov}
\affiliation{Petersburg Nuclear Physics Institute, NRC Kurchatov Institute, 188300 Gatchina, Russia}
\author{V.\,N.\,Fedosseev}
\affiliation{CERN, CH-1211 Geneva 23, Switzerland}
\author{G.\,Fern\'andez-Mart\'inez}
\affiliation{Institut f\"ur Kernphysik, Technische Universit\"at zu Darmstadt, 64289 Darmstadt, Germany}
\author{A.\,Fija\l{}kowska}
\affiliation{Faculty of Physics, University of Warsaw, PL 02-093 Warsaw, Poland}
\author{H.\,Fynbo}
\affiliation{Department of Physics and Astronomy, Aarhus University, DK-8000 Aarhus C, Denmark}
\author{D.\,Galaviz}
\affiliation{LIP, and Faculty of Sciences, University of Lisbon, 1000-149 Lisbon, Portugal}
\author{P. Galve}
\affiliation{Grupo de F\'isica Nuclear and IPARCOS, Universidad Complutense de Madrid, CEI Moncloa, E-28040 Madrid, Spain}
%
 \author{M.\,Garc\'{i}a-D\'{i}ez}
\affiliation{Grupo de F\'isica Nuclear and IPARCOS, Universidad Complutense de Madrid, CEI Moncloa, E-28040 Madrid, Spain}
\author{P.\,T.\,Greenlees}
\affiliation{University of Jyv\"{a}skyl\"{a}, Department of Physics, P.O. Box 35, FI-40014, Jyv\"{a}skyl\"{a}, Finland}
\affiliation{Helsinki Institute of Physics, University of Helsinki, FIN-00014 Helsinki, Finland}
\author{R.\,Grzywacz}
\affiliation{Department of Physics and Astronomy, University of Tennessee, Knoxville, Tennessee 37996, USA}
\affiliation{Physics Division, Oak Ridge National Laboratory, Oak Ridge, Tennessee 37831, USA}
\author{L.\,J.\,Harkness-Brennan}
\affiliation{Oliver Lodge Laboratory, The University of Liverpool, Liverpool, L69 7ZE, United Kingdom}
\author{C.\,Henrich}
\affiliation{Institut f\"ur Kernphysik, Technische Universit\"at Darmstadt, 64289 Darmstadt, Germany}
\author{M.\,Huyse}
\affiliation{KU Leuven, Instituut voor Kern- en Stralingsfysica, Celestijnenlaan 200D, 3001 Leuven, Belgium}
\author{P. Ib\'{a}\~{n}ez}
\affiliation{Grupo de F\'isica Nuclear and IPARCOS, Universidad Complutense de Madrid, CEI Moncloa, E-28040 Madrid, Spain}
\author{A.\,Illana}
\affiliation{KU Leuven, Instituut voor Kern- en Stralingsfysica, Celestijnenlaan 200D, 3001 Leuven, Belgium}
\affiliation{Instituto Nazionale di Fisica Nucleare, Laboratori Nazionali di Legnaro, I-35020 Legnaro, Italy}
\author{Z.\,Janas}
\affiliation{Faculty of Physics, University of Warsaw, PL 02-093 Warsaw, Poland}
\author{K.\,Johnston}
\affiliation{CERN, CH-1211 Geneva 23, Switzerland}
\author{J.\,Jolie}
\affiliation{Institut f\"ur Kernphysik, Universit\"at zu K\"oln, 50937 K\"oln, Germany}
\author{D.\,S.\,Judson}
\affiliation{Oliver Lodge Laboratory, The University of Liverpool, Liverpool, L69 7ZE, United Kingdom}
\author{V.\,Karanyonchev}
\affiliation{Institut f\"ur Kernphysik, Universit\"at zu K\"oln, 50937 K\"oln, Germany}
\author{M.\,Kici\'nska-Habior}
\affiliation{Faculty of Physics, University of Warsaw, PL 02-093 Warsaw, Poland}
\author{J.\,Konki}
\affiliation{University of Jyv\"{a}skyl\"{a}, Department of Physics, P.O. Box 35, FI-40014, Jyv\"{a}skyl\"{a}, Finland}
\affiliation{Helsinki Institute of Physics, University of Helsinki, FIN-00014 Helsinki, Finland}
\author{\L{}.\,Koszuk}
\affiliation{Faculty of Physics, University of Warsaw, PL 02-093 Warsaw, Poland}
\author{J.\,Kurcewicz}
\affiliation{CERN, CH-1211 Geneva 23, Switzerland}
\author{I.\,Lazarus}
\affiliation{STFC Daresbury, Daresbury, Warrington WA4 4AD, United Kingdom}
\author{R.\,Lic\u{a}}
\affiliation{CERN, CH-1211 Geneva 23, Switzerland}
\affiliation{``Horia Hulubei'' National Institute of Physics and Nuclear Engineering, RO-077125 Bucharest, Romania}
\author{A. L\'{o}pez-Montes}
\affiliation{Grupo de F\'isica Nuclear and IPARCOS, Universidad Complutense de Madrid, CEI Moncloa, E-28040 Madrid, Spain}
%
\author{H.\,Mach}
\affiliation{National Centre for Nuclear Research, BP1, PL 00-681 Warsaw, Poland}
\author{M.\,Madurga}
\affiliation{CERN, CH-1211 Geneva 23, Switzerland}
\affiliation{Department of Physics and Astronomy, University of Tennessee, Knoxville, Tennessee 37996, USA}
\author{I.\,Marroqu\'in}
\affiliation{Instituto de Estructura de la Materia, CSIC, E-28006 Madrid, Spain}
\author{B.\,Marsh}
\affiliation{CERN, CH-1211 Geneva 23, Switzerland}
\author{M.\,C.\,Mart\'inez}
\affiliation{Grupo de F\'isica Nuclear and IPARCOS, Universidad Complutense de Madrid, CEI Moncloa, E-28040 Madrid, Spain}
\author{C.\,Mazzocchi}
\affiliation{Faculty of Physics, University of Warsaw, PL 02-093 Warsaw, Poland}
\author{K.\,Miernik}
\affiliation{Faculty of Physics, University of Warsaw, PL 02-093 Warsaw, Poland}
\author{C.\,Mihai}
\affiliation{``Horia Hulubei'' National Institute of Physics and Nuclear Engineering, RO-077125 Bucharest, Romania}
\author{N. M\u{a}rginean}
\affiliation{``Horia Hulubei'' National Institute of Physics and Nuclear Engineering, RO-077125 Bucharest, Romania}
\author{R. M\u{a}rginean}
\affiliation{``Horia Hulubei'' National Institute of Physics and Nuclear Engineering, RO-077125 Bucharest, Romania}
\author{A.\,Negret}
\affiliation{``Horia Hulubei'' National Institute of Physics and Nuclear Engineering, RO-077125 Bucharest, Romania}
\author{E.\,N\'acher}
\affiliation{Instituto de F\'isica Corpuscular, CSIC-Universidad de Valencia, E-46071 Valencia, Spain}
\author{J. Ojala}
\affiliation{University of Jyv\"{a}skyl\"{a}, Department of Physics, P.O. Box 35, FI-40014, Jyv\"{a}skyl\"{a}, Finland}
\author{B.\,Olaizola}
\affiliation{Department of Physics, University of Guelph, Guelph,
Ontario, Canada N1G 2W1}
\affiliation{TRIUMF, 4004 Wesbrook Mall, Vancouver, British Columbia V6T
2A3, Canada}
\affiliation{CERN, CH-1211 Geneva 23, Switzerland}
\author{R.\,D.\,Page}
\affiliation{Oliver Lodge Laboratory, The University of Liverpool, Liverpool, L69 7ZE, United Kingdom}
\author{J. Pakarinen}
\affiliation{University of Jyv\"{a}skyl\"{a}, Department of Physics, P.O. Box 35, FI-40014, Jyv\"{a}skyl\"{a}, Finland}
\author{S.\,Pascu}
\affiliation{``Horia Hulubei'' National Institute of Physics and Nuclear Engineering, RO-077125 Bucharest, Romania}
\author{S.\,V.\,Paulauskas}
\affiliation{Department of Physics and Astronomy, University of Tennessee, Knoxville, Tennessee 37996, USA}
\author{A.\,Perea}
\affiliation{Instituto de Estructura de la Materia, CSIC, E-28006 Madrid, Spain}
\author{V.\,Pucknell}
\affiliation{STFC Daresbury, Daresbury, Warrington WA4 4AD, United Kingdom}
\author{P.\,Rahkila}
\affiliation{University of Jyv\"{a}skyl\"{a}, Department of Physics, P.O. Box 35, FI-40014, Jyv\"{a}skyl\"{a}, Finland}
\affiliation{Helsinki Institute of Physics, University of Helsinki, FIN-00014 Helsinki, Finland}
\author{C.\,Raison}
\affiliation{Department of Physics, University of York, York, YO10 5DD, United Kingdom}
%
\author{E.\,Rapisarda}
\affiliation{CERN, CH-1211 Geneva 23, Switzerland}
\author{K.\,Rezynkina}
\affiliation{KU Leuven, Instituut voor Kern- en Stralingsfysica, Celestijnenlaan 200D, 3001 Leuven, Belgium}
\author{F.\,Rotaru}
\affiliation{``Horia Hulubei'' National Institute of Physics and Nuclear Engineering, RO-077125 Bucharest, Romania}
\author{S.\,Rothe}
\affiliation{CERN, CH-1211 Geneva 23, Switzerland}
\author{K.\,P.\,Rykaczewski}
\affiliation{Physics Division, Oak Ridge National Laboratory, Oak Ridge, Tennessee 37831, USA}
\author{J.-M.\,R\'egis}
\affiliation{Institut f\"ur Kernphysik, Universit\"at zu K\"oln, 50937 K\"oln, Germany}
\author{K.\,Schomacker}
\affiliation{Institut f\"ur Kernphysik, Universit\"at zu K\"oln, 50937 K\"oln, Germany}
%
\author{M.\,Si\l{}kowski}
\affiliation{Faculty of Physics, University of Warsaw, PL 02-093 Warsaw, Poland}
\author{G.\,Simpson}
\affiliation{Laboratoire de Physique Subatomique et de Cosmologie, IN2P3-CNRS/Universit\'e Grenoble Alpes, Grenoble Cedex F-38026, France}
\author{C.\,Sotty}
\affiliation{``Horia Hulubei'' National Institute of Physics and Nuclear Engineering, RO-077125 Bucharest, Romania}
\affiliation{KU Leuven, Instituut voor Kern- en Stralingsfysica, Celestijnenlaan 200D, 3001 Leuven, Belgium}
\author{L.\,Stan}
\affiliation{``Horia Hulubei'' National Institute of Physics and Nuclear Engineering, RO-077125 Bucharest, Romania}
\author{M.\,St\u{a}noiu}
\affiliation{``Horia Hulubei'' National Institute of Physics and Nuclear Engineering, RO-077125 Bucharest, Romania}
\author{M.\,Stryjczyk}
\affiliation{Faculty of Physics, University of Warsaw, PL 02-093 Warsaw, Poland}
\affiliation{KU Leuven, Instituut voor Kern- en Stralingsfysica, Celestijnenlaan 200D, 3001 Leuven, Belgium}
\affiliation{University of Jyv\"{a}skyl\"{a}, Department of Physics, P.O. Box 35, FI-40014, Jyv\"{a}skyl\"{a}, Finland}
\author{D.\,S\'{a}nchez-Parcerisa}
\affiliation{Grupo de F\'isica Nuclear and IPARCOS, Universidad Complutense de Madrid, CEI Moncloa, E-28040 Madrid, Spain}
%
\author{V.\,S\'anchez-Tembleque}
\affiliation{Grupo de F\'isica Nuclear and IPARCOS, Universidad Complutense de Madrid, CEI Moncloa, E-28040 Madrid, Spain}
\author{O.\,Tengblad}
\affiliation{Instituto de Estructura de la Materia, CSIC, E-28006 Madrid, Spain}
\author{A.\,Turturic\u{a}}
\affiliation{``Horia Hulubei'' National Institute of Physics and Nuclear Engineering, RO-077125 Bucharest, Romania}
\author{J.\,M.\,Ud\'ias}
\affiliation{Grupo de F\'isica Nuclear and IPARCOS, Universidad Complutense de Madrid, CEI Moncloa, E-28040 Madrid, Spain}
\author{P.\,Van~Duppen}
\affiliation{KU Leuven, Instituut voor Kern- en Stralingsfysica, Celestijnenlaan 200D, 3001 Leuven, Belgium}
\author{V.\,Vedia}
\affiliation{Grupo de F\'isica Nuclear and IPARCOS, Universidad Complutense de Madrid, CEI Moncloa, E-28040 Madrid, Spain}
\author{A.\,Villa}
\affiliation{Grupo de F\'isica Nuclear and IPARCOS, Universidad Complutense de Madrid, CEI Moncloa, E-28040 Madrid, Spain}
\author{S. Vi\~{n}als}
\affiliation{Instituto de Estructura de la Materia, CSIC, E-28006 Madrid, Spain}
\author{R.\,Wadsworth}
\affiliation{Department of Physics, University of York, York, YO10 5DD, United Kingdom}
\author{W.\,B.\,Walters}
\affiliation{Department of Chemistry, University of Maryland, Maryland 20742, USA}
\author{N.\,Warr}
\affiliation{Institut f\"ur Kernphysik, Universit\"at zu K\"oln, 50937 K\"oln, Germany}
\author{S.\,G.\,Wilkins}
\affiliation{CERN, CH-1211 Geneva 23, Switzerland}
\collaboration{IDS Collaboration}
\noaffiliation
\date{\today}
%
\begin{abstract}
%
%
The $\beta$ decay of the neutron-rich \textsuperscript{134}In and \textsuperscript{135}In was investigated experimentally in~order to provide new insights into the nuclear structure of the tin isotopes with magic proton number $Z=50$ above the $N=82$ shell.
The~$\beta$-delayed $\gamma$-ray spectroscopy measurement was performed at~the ISOLDE facility at CERN, where indium isotopes were selectively laser-ionized and on-line mass separated. 
Three $\beta$-decay branches of \textsuperscript{134}In were established, two of which were observed for the first time. Population of neutron-unbound states decaying via $\gamma$~rays was identified in~the two daughter nuclei of~\textsuperscript{134}In, \textsuperscript{134}Sn and~\textsuperscript{133}Sn, at~excitation energies exceeding the neutron separation energy by~1~MeV. The $\beta$-delayed one- and two-neutron emission branching ratios of~\textsuperscript{134}In were determined and~compared with theoretical calculations. The~$\beta$-delayed one-neutron decay was observed to~be~dominant $\beta$-decay branch of~\textsuperscript{134}In even though the Gamow-Teller resonance is~located substantially above the two-neutron separation energy of~\textsuperscript{134}Sn. 
%
Transitions following the $\beta$~decay of~\textsuperscript{135}In are reported for the first time, including $\gamma$~rays tentatively attributed to~\textsuperscript{135}Sn. 
In~total, six new levels were identified in~\textsuperscript{134}Sn on the basis of the $\beta \gamma \gamma$ coincidences observed in~the \textsuperscript{134}In and \textsuperscript{135}In $\beta$~decays. A~transition that might be a~candidate for deexciting the missing neutron single-particle $13/2^+$ state in~\textsuperscript{133}Sn was observed in~both $\beta$~decays and~its assignment is~discussed. Experimental level schemes of \textsuperscript{134}Sn and \textsuperscript{135}Sn are compared with shell-model predictions.
Using the fast timing technique, half-lives~of~the $2^+$, $4^+$ and $6^+$~levels in~\textsuperscript{134}Sn were~determined. From~the lifetime of~the $4^+$ state measured for~the first time, an~unexpectedly large B($E2$; $4^+\rightarrow 2^+$) transition strength was~deduced, which is~not reproduced by~the shell-model calculations. 
%
%
\end{abstract}

\maketitle
\section{Introduction}
The region around \textsuperscript{132}Sn, the heaviest doubly-magic nucleus far from the valley of $\beta$-stability, is~of great relevance for the development of the theoretical description of neutron-rich nuclei. 
New experimental data for nuclei in~that region allow for a~better understanding of phenomena that occur when the $N/Z$ ratio becomes large, such as evolution of shell structure~\cite{10.1103/PhysRevLett.95.232502, 10.1103/RevModPhys.92.015002, 10.1016/j.ppnp.2008.05.001, 10.1103/PhysRevLett.122.212502} and rare processes of $\beta$-delayed multiple-neutron emission~\cite{10.1016/j.nds.2015.08.002, 10.1016/j.nds.2020.09.001, 10.1103/PhysRevC.93.025805, 10.1103/PhysRevC.90.054306}.
Properties of nuclei around \textsuperscript{132}Sn are also important for modeling the rapid neutron capture nucleosynthesis process (\emph{r}-process), since the $A \approx 130$ peak in~the \emph{r}-process abundance pattern is linked to the $N=82$ shell closure~\cite{10.1103/RevModPhys.29.547, 10.1140/epjad/i2005-06-157-2, 10.1016/S0375-9474(01)01141-1, 10.1016/j.ppnp.2015.09.001}. \par
Due to the robust nature of the \textsuperscript{132}Sn core~\cite{10.1038/nature09048}, tin isotopes above $N=82$ offer a~rare opportunity to investigate neutron-neutron components of effective nucleon-nucleon interactions for heavy-mass nuclei with large neutron excess~\cite{10.1103/PhysRevLett.113.132502}.
At~present, the \textsuperscript{132}Sn region is~a~unique part of the chart of nuclides where spectroscopic information for neutron-rich nuclei with one and few neutrons beyond the double-shell closure was obtained~\cite{10.1103/PhysRevLett.77.1020, 10.1007/s002180050269, 10.1103/PhysRevLett.113.132502}. 
The \textsuperscript{133}Sn nucleus, with only one neutron outside the doubly magic \textsuperscript{132}Sn, is~the heaviest odd-\emph{A} tin isotope for which excited states were reported so~far~\cite{10.1103/PhysRevLett.77.1020, 10.1023/A:1012661816716, 10.1038/nature09048, 10.1103/PhysRevC.84.034601, 10.1103/PhysRevLett.112.172701, 10.1103/PhysRevLett.118.202502, 10.5506/APhysPolB.49.523, 10.1103/PhysRevC.99.024304}. This nuclide has been extensively studied for over two decades to gain information about neutron~($\nu$) single-particle (s.\,p.) states just outside the closed shell at $N=82$. Still, the energy of the $\nu1i_{13/2}$ s.\,p.~state in~\textsuperscript{133}Sn remains unknown.
Recently, states having dominant two-particle one-hole (2\emph{p}1\emph{h}) neutron configurations with respect to the \textsuperscript{132}Sn core were identified in~\textsuperscript{133}Sn~\cite{10.1103/PhysRevLett.118.202502, 10.1103/PhysRevC.99.024304}. 
In~the case of even-\emph{A} tin isotopes above $N=82$, information on excited states was obtained for \textsuperscript{134}Sn, \textsuperscript{136}Sn and \textsuperscript{138}Sn~\cite{10.1007/s002180050269, 10.1007/PL00013594, 10.1103/PhysRevLett.113.132502, 10.1103/PhysRevC.86.054319}. All~members of the two-neutron $\nu 2f_{7/2}$ ($\nu 2f_{7/2}^{\,2}$) multiplet were reported in~these isotopes. An~additional state belonging to the $\nu 2f_{7/2} 1h_{9/2}$ configuration is~known in~\textsuperscript{134}Sn~\cite{10.1007/PL00013594}. 
Despite extensive studies, information on~tin isotopes beyond $N=82$ still appears to be scarce. \par
In the present work, we report on the results of a~$\beta$-decay study of \textsuperscript{134}In and \textsuperscript{135}In nuclei that provide new experimental insights into tin isotopes above $N=82$. 
In~an~\emph{r}-process sensitivity study, \textsuperscript{134}In and \textsuperscript{135}In were indicated to be among those $\beta$-delayed neutron ($\beta n$) emitters that have the greatest impact on the abundance pattern in~cold wind \emph{r}-process simulations~\cite{10.7566/JPSCP.6.010010}. Moreover, for neutron densities around $10^{25}$\,cm\textsuperscript{-3}, where the \emph{r}-matter flow has already broken through the $N= 82$ shell, the \textsuperscript{135}In nuclide acts as an important waiting-point~\cite{10.1007/s10050-002-8756-7}. \par
The neutron-rich isotopes \textsuperscript{134}In and \textsuperscript{135}In represent rare cases of experimentally accessible nuclei for which the $\beta$-delayed three neutron ($\beta 3n$) decay is~energetically allowed~\cite{10.1016/j.nds.2020.09.001, 10.1088/1674-1137/abddaf}.
%
Therefore, these isotopes are representative nuclei to investigate competition between $\beta$-delayed one- ($\beta 1n$) and multiple-neutron ($\beta 2n$, $\beta 3n$,\,...) emission as well as the $\gamma$-ray contribution to the decay of~neutron-unbound states~\cite{10.1103/PhysRevC.94.064317, 10.1103/PhysRevC.100.031302}. 
Recently, a~significant $\gamma$-ray branch for levels above the neutron separation energy ($S_n$) was observed in~\textsuperscript{133}Sn~\cite{10.1103/PhysRevLett.118.202502, 10.1103/PhysRevC.99.024304}.
%
%
As~reported in~Ref.~\cite{10.1103/PhysRevLett.118.202502}, the main factor that makes the neutron emission from highly excited 2\emph{p}1\emph{h} states in~\textsuperscript{133}Sn hindered is~the small overlap of the wave functions of the states involved in~the $\beta n$ decay. It~is expected that similar nuclear structure effects play a~role for other nuclei southeast of \textsuperscript{132}Sn, including \textsuperscript{134}In and \textsuperscript{135}In~\cite{10.1103/PhysRevLett.118.202502}.\par
%
%
So far, the $\beta$ decay of \textsuperscript{134}In was investigated via $\beta$-delayed $\gamma$-ray spectroscopy in~only one measurement, which provided the first information about neutron s.\,p.~states in \textsuperscript{133}Sn~\cite{10.1103/PhysRevLett.77.1020, 10.1023/A:1012661816716}. 
The population of excited states in~other tin isotopes was not observed. The~$\beta n$~emission probability~($P_n$) was estimated to~be~around $65\%$ and a~$\beta$-decay half-life of 138(8)~ms was reported for \textsuperscript{134}In~\cite{10.1103/PhysRevLett.77.1020, 10.1016/j.nds.2020.09.001}. Later, the measurement of $\beta$-delayed neutrons from \textsuperscript{134}In yielded the~more precise value of 141(5)~ms~\cite{10.1007/s10050-002-8756-7}. Recently, a~half-life of 126(7)~ms was obtained at RIKEN for \textsuperscript{134}In~\cite{10.1103/PhysRevLett.114.192501}.
In~the case of the \textsuperscript{135}In $\beta$ decay, no~information on~the population of states in~tin isotopes existed prior to this work. The~$\beta$-decay half-life of~\textsuperscript{135}In was measured in two experiments, which yielded values of~92(10)~ms~\cite{10.1007/s10050-002-8756-7} and~103(5)~ms~\cite{10.1103/PhysRevLett.114.192501}, respectively. \par
In~this work, we observed for the first time the $\beta$-decay ($\beta \gamma$) and $\beta 2n$-decay branches of \textsuperscript{134}In. Transitions following the \textsuperscript{135}In $\beta$ decay, including those belonging to the $\beta \gamma$-, $\beta 1n$- and $\beta 2n$-decay branches, were also established for the first time. \par
\par
\section{Experimental details}
The \textsuperscript{134}In and \textsuperscript{135}In nuclei were produced at the ISOLDE-CERN facility~\cite{10.1088/1361-6471/aa7eba}. The~1.4-GeV proton beam from the Proton Synchrotron Booster (PSB) was directed onto a~solid tungsten proton-to-neutron converter~\cite{10.1016/j.nimb.2014.04.026} producing spallation neutrons that induced fission in a~thick uranium carbide target. The indium atoms diffused out of the target material and subsequently effused via a~transfer line into the hot cavity ion source, where they were selectively ionized by the Resonance Ionization Laser Ion Source (RILIS)~\cite{10.1088/1361-6471/aa78e0}. 
After extraction and acceleration by a~40\,kV potential, the indium isotopes were separated according to the mass-to-charge ratio by the General Purpose Separator and then transmitted to the ISOLDE Decay Station (IDS)~\cite{IDS-web}. They were implanted on an aluminized mylar tape at the center of~the detection setup. The~time structure of~ions reaching IDS varied depending on~the composition of a~repetitive sequence of~proton pulses, called the supercycle, distributed by~the PSB at~intervals of 1.2\,s. 
The~supercycle structure varied during the~experiment and its~length ranged from 26 to~34 proton pulses, corresponding to~31.2~and~40.8~s, respectively. The~extraction of~the ion beam was started 5~ms after each proton pulse from~PSB and lasted 500~ms for \textsuperscript{134}In and 225~ms for \textsuperscript{135}In. 
\par
%
Data were collected during the beam implantation and the subsequent decay of~the isotopes of~interest. 
%
%
To~suppress the long-lived activity from the decay of daughter nuclei, the tape was moved after each supercycle.
Additional measurements were performed with the \textsuperscript{134}In beam in~which the tape was moved after each proton pulse. 
Surface-ionized isobaric contaminants, \textsuperscript{134}Cs and \textsuperscript{135}Cs, were present in~the $A=134$ and $A=135$ ion beams, respectively. 
In~the case of the $A=135$ measurements, the~isomeric state of \textsuperscript{135}Cs was a~severe source of background. 
For identification of beam impurities, an~additional measurement was performed at mass $A=135$ with one of the RILIS lasers turned off. In~such laser-off mode, only surface-ionized elements reached the IDS, while in~the laser-on mode, RILIS-ionized indium was additionally present in the beam. \par
%
%
To detect $\beta$ particles, a~fast-response 3-mm-thick NE111A plastic scintillator was used. It was positioned directly behind the ion collection point and provided a~detection efficiency of around 20\%. For the $\gamma$-ray detection, four high-purity germanium (HPGe) Clover-type detectors and two truncated cone-shaped LaBr$_3$(Ce) crystals~\cite{10.1016/j.nima.2017.03.030} coupled to fast photomultiplier tubes (PMTs) were utilized.  The PMT anode signals from fast-response detectors were processed by analog constant fraction discriminators and then sent to time-to-amplitude converters (TACs), which provided the time difference between coincident signals from plastic and~LaBr$_3$(Ce) detectors. With~this configuration, it~was possible to~perform lifetime measurements for~excited states using the advanced time-delayed $\beta \gamma \gamma$(t) (fast timing) technique~\cite{10.1016/0168-9002(89)91272-2, 10.1016/0168-9002(89)90770-5, 10.1088/1361-6471/aa8217}. \par
The Nutaq digital data acquisition system~\cite{nutaq-web} was used to record and sample energy signals from all detectors along with outputs from TACs and reference signal from the PSB. Data were collected in a~triggerless mode. Events were reconstructed in~the offline analysis, in which they were correlated with the occurrence of the proton pulse. \par
Energy and efficiency calibrations of $\gamma$-ray detectors were performed using \textsuperscript{152}Eu, \textsuperscript{140}Ba-\textsuperscript{140}La and \textsuperscript{133}Ba radioactive sources as well as \textsuperscript{88}Rb and \textsuperscript{138}Cs samples produced on-line. High-energy $\gamma$~rays originating from the background induced by neutrons from the target area were used to extend the energy calibration of HPGe detectors up to 7.6~MeV. The $\gamma$-ray photopeak efficiency of the HPGe detectors reached 4\% at 1173~keV after the add-back procedure. For each LaBr$_3$(Ce) detector, an~efficiency of around 1\% at 1~MeV was obtained. Time-response calibrations of LaBr$_3$(Ce) detectors for full-energy peaks as a~function of $\gamma$-ray energy as well as corrections due to Compton events were included in the fast-timing analysis. More details on the lifetime measurements using the same experimental setup are provided in Refs.~\cite{10.1103/PhysRevC.102.014328, 10.1088/1361-6471/aa8217, 10.1088/1361-6471/aa6015, JB-thesis}.
\section{Results}
\subsection{\protect\boldmath$\beta$ decay of \textsuperscript{134}In}
\begin{figure*}
\includegraphics[width=\linewidth]{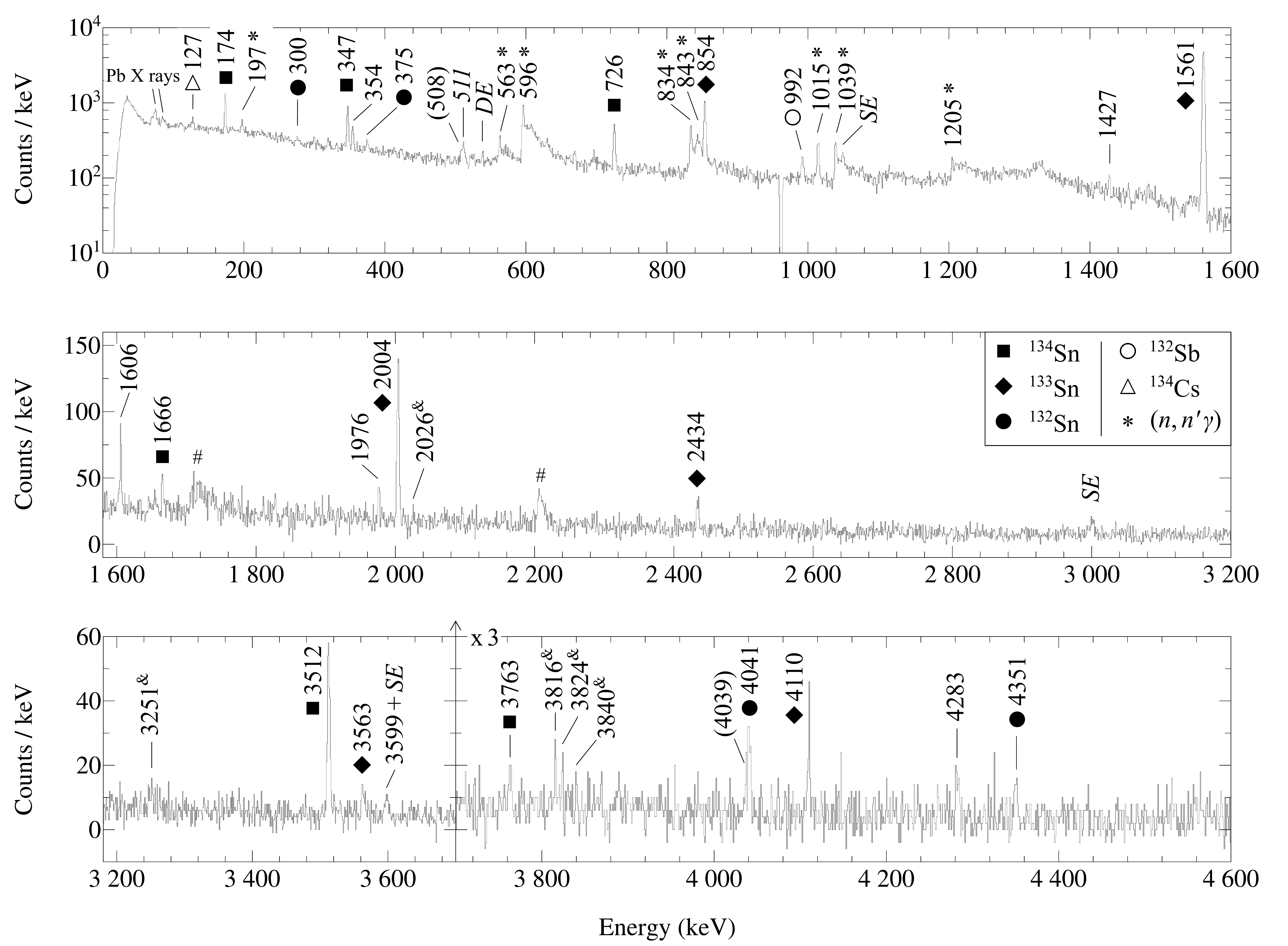}
\caption{
The $\beta$-gated $\gamma$-ray spectrum obtained at $A=134$ in~the first 400~ms relative to the~proton pulse from~which long-lived background has been subtracted. 
Transitions assigned to the daughter nuclei of \textsuperscript{134}In are labeled with filled symbols, while those attributed to activities of daughter or contaminant nuclei are marked with open symbols. 
Transitions that can be assigned to the \textsuperscript{134}In $\beta$ decay but not to a specific decay branch are indicated by energy only. 
Lines marked with an~ampersand indicate possible weak transitions whose identification is~uncertain.
Energies of~possible peaks, which might correspond to~artifacts due~to~the background subtraction procedure, are given in~parentheses. 
The presence of a~negative peak at 962 keV is~the consequence of subtracting the contribution from the daughter nucleus \textsuperscript{133}Sn~\cite{10.1007/BF01879878}.
Triangular-shaped peaks arising from inelastic neutron scattering in the HPGe detectors~\cite{10.1016/j.nima.2004.11.021, 10.1140/epja/i2007-10553-8, 10.1140/epja/i2013-13028-5, 10.1016/0969-8043(93)90177-C, 10.1016/j.apradiso.2020.109422} are indicated with asterisks. The peak at 197~keV is~also considered as induced by neutrons~\cite{10.1140/epja/i2013-13028-5}. The abbreviations \emph{SE} and \emph{DE} indicate single-escape and double-escape peaks, respectively. Broad peaks marked with a~hash symbol remained unidentified. 
}
\label{fig:134In_beta-gated_spectrum}
\end{figure*}

Transitions following the $\beta$ decay of \textsuperscript{134}In were identified by~comparing $\beta$-gated $\gamma$-ray spectra sorted using various conditions on~the time of~the event with respect to~the proton pulse. Lines that can be attributed to~$\gamma$~rays in~daughter nuclei are~enhanced when~this time window is~limited to a~few hundred milliseconds. 
Figure~\ref{fig:134In_beta-gated_spectrum} shows the $\beta$-gated $\gamma$-ray spectrum obtained at~$A=134$ during the first 400~ms following the proton pulse. Long-lived background, originating from decays of daughter nuclei and the surface-ionized \textsuperscript{134$m$}Cs contaminant, was subtracted from the data presented.
Apart from $\gamma$~rays that can be assigned to the \textsuperscript{134}In $\beta$~decay, neutron-induced background arising from inelastic scattering of~fast neutrons~\cite{10.1016/j.nima.2004.11.021, 10.1140/epja/i2007-10553-8, 10.1140/epja/i2013-13028-5, 10.1016/0969-8043(93)90177-C, 10.1016/j.apradiso.2020.109422}, which were emitted from~\textsuperscript{134}In as~$\beta$-delayed particles, is~also prominent. \par
\begin{figure}
\includegraphics[width=\columnwidth]{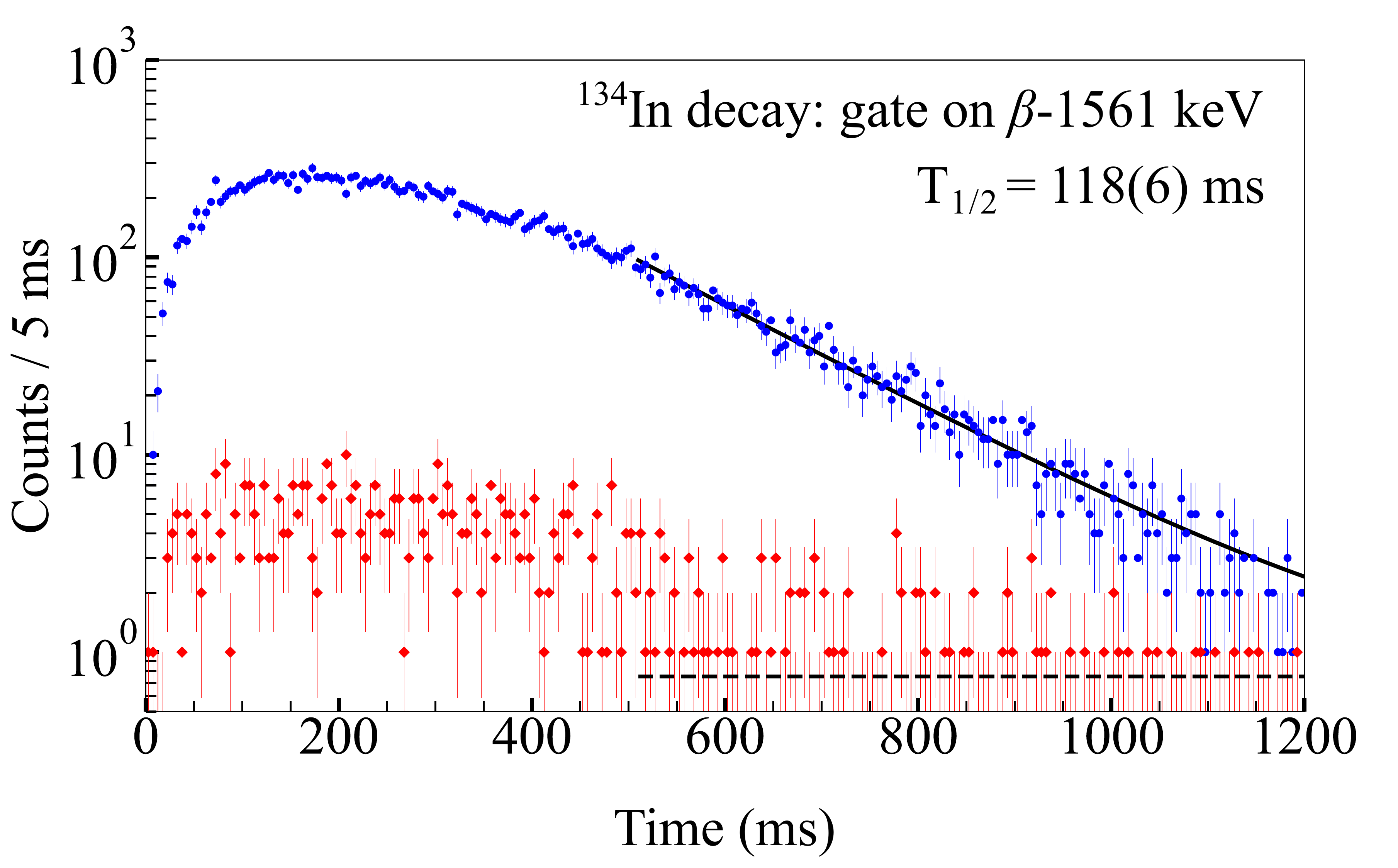}
\caption{(Color online) Time distributions relative to the proton pulse of the 1561-keV transition (blue circles) and the background area (red diamonds) observed in~coincidence with $\beta$~particles at $A=134$ when RILIS was applied to ionize indium. A~function composed of an exponential decay and a~constant background was fitted (solid line) in~the 510-1200~ms time range. The curve corresponding to the background component is~also presented (dashed line). A~Bayesian approach was applied in the statistical analysis of the data~\cite{10.1103/PhysRevC.98.064326}. }
\label{fig:134In_T1-2_1561keV}
\end{figure}

The three most intense lines in~the spectrum shown in~Fig.~\ref{fig:134In_beta-gated_spectrum}, at energies of 854, 1561 and 2004~keV, were observed in~the previous $\beta$-decay study of \textsuperscript{134}In~\cite{10.1103/PhysRevLett.77.1020, 10.1023/A:1012661816716}. They were assigned to the \textsuperscript{133}Sn nucleus as transitions depopulating the $3/2^-$, $(9/2^-)$ and $5/2^-$ states, respectively. These assignments were confirmed later in~one-neutron transfer reactions~\cite{10.1038/nature09048, 10.1103/PhysRevC.84.034601, 10.1103/PhysRevLett.112.172701}. 
The most intense transition at 1561~keV was used to determine the $\beta$-decay half-life of \textsuperscript{134}In. From the time distribution relative to the proton pulse, shown in Fig.~\ref{fig:134In_T1-2_1561keV}, the~half-life was deduced to be 118(6)~ms.
This value is~consistent with the \textsuperscript{134}In half-life measured recently at RIKEN, 126(7)~ms~\cite{10.1103/PhysRevLett.114.192501} and slightly differs from the values previously reported in~Ref.~\cite{10.1103/PhysRevLett.77.1020}, 138(8)~ms, and in~Ref.~\cite{10.1007/s10050-002-8756-7}, 141(5)~ms. \par
In the present work, the $\beta 1n$-decay branch of \textsuperscript{134}In is~expanded with three transitions, all of which depopulate states above $S_n$ in~\textsuperscript{133}Sn, 2398.7(27)~keV~\cite{10.1088/1674-1137/abddaf}. The peak visible in Fig.~\ref{fig:134In_beta-gated_spectrum} at 3563~keV corresponds to the transition depopulating the $(11/2^-)$ state in \textsuperscript{133}Sn. A~3570(50)-keV $\gamma$~ray was first identified in~\textsuperscript{133}Sn via one-neutron knockout from \textsuperscript{134}Sn~\cite{10.1103/PhysRevLett.118.202502}. This was confirmed in~a~$\beta$-decay study of \textsuperscript{133}In that provided improved precision of its energy, 3563.9(5)~keV~\cite{10.1103/PhysRevC.99.024304}.
The peak visible in~Fig.~\ref{fig:134In_beta-gated_spectrum} at 4110~keV can be associated with the 4110.8(3)-keV $\gamma$~ray, which was seen previously in~the $\beta$~decay of \textsuperscript{133}In~\cite{10.1103/PhysRevC.102.014328}, but the absence of $\beta \gamma \gamma$ coincidence relations hindered its assignment to a~particular daughter nucleus. An~observation of this line in~the $\beta$ decays of both \textsuperscript{133}In and \textsuperscript{134}In provides support for its assignment to the \textsuperscript{133}Sn nucleus. \par
%
In~the energy range corresponding to the predicted excitation energy of the $13/2^+$ state in~\textsuperscript{133}Sn, 2511(80)~keV~\cite{10.1103/PhysRevC.91.027303} or~between 2360 and~2600~keV~\cite{10.1103/PhysRevC.94.034309}, one~relatively-intense transition was registered at~2434~keV (see~Fig.~\ref{fig:134In_beta-gated_spectrum}). No~$\beta \gamma \gamma$ and $\gamma \gamma$~coincidence relationships were observed for this line, making its assignment to either \textsuperscript{134}Sn or \textsuperscript{132}Sn unlikely and thus providing an argument for its assignment to~\textsuperscript{133}Sn. The~2792-keV transition, discussed in~Ref.~\cite{10.1103/PhysRevLett.112.172701} as~a possible candidate for $\gamma$~ray depopulating the $13/2^+$ state in~ \textsuperscript{133}Sn, was not observed in~the $\beta$~decay of \textsuperscript{134}In. 
\par
Among~the known low-lying levels in~\textsuperscript{133}Sn, only~the $1/2^-$~state~\cite{10.1038/nature09048, 10.1103/PhysRevLett.112.172701, 10.1103/PhysRevC.99.024304} was~not seen in~the \textsuperscript{134}In $\beta$~decay. The 354-keV transition that was identified in~the previous $\beta$-decay study of \textsuperscript{134}In but remained unassigned despite being registered in~coincidence with $\beta$-delayed neutrons~\cite{10.1103/PhysRevLett.77.1020, 10.1023/A:1012661816716}, was~observed in~the present study. No~$\beta \gamma \gamma$ and $\gamma \gamma$ coincidence relations were found for this transition, making its attribution to any of the daughter nuclei impossible. %
The 802-keV transition for which a~coincidence with neutrons emitted from \textsuperscript{134}In was also reported in~Refs.~\cite{10.1103/PhysRevLett.77.1020, 10.1023/A:1012661816716} was~not present in~our spectra.\par
We now turn to the $\beta \gamma$-decay branch of \textsuperscript{134}In, leading to the population of states in~\textsuperscript{134}Sn, which was observed for the first time in this work. Figure~\ref{fig:134In_beta-gated_spectrum} shows clearly the presence of the 174-, 347- and 726-keV transitions that were assigned to the yrast $6^+ \rightarrow 4^+ \rightarrow 2^+ \rightarrow0^+_{g.s.}$ cascade in~\textsuperscript{134}Sn from the \textsuperscript{248}Cm fission data~\cite{10.1007/s002180050269, 10.1007/PL00013594}. 
The~1262-keV $\gamma$~ray deexciting the $(8^+)$~state in~\textsuperscript{134}Sn~\cite{10.1007/PL00013594} was~not observed in~the \textsuperscript{134}In~$\beta$~decay. \par
\begin{figure}
\includegraphics[width=\linewidth]{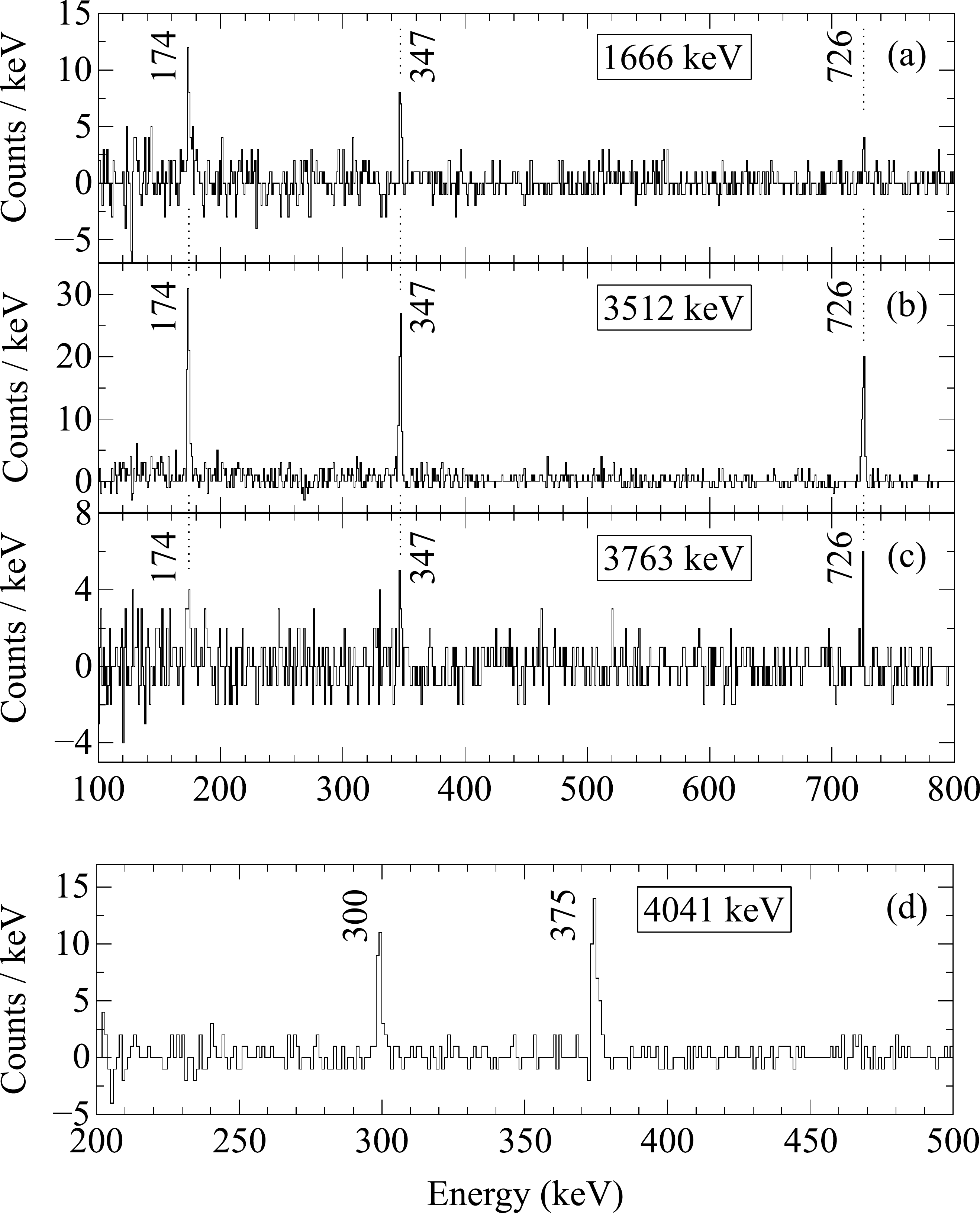}
\caption{Background-subtracted $\gamma$-ray spectra in coincidence with the (a)\,1666-, (b)\,3512- and (c) 3763-keV transitions that depopulate new levels in~\textsuperscript{134}Sn following the $\beta \gamma$~decay of~\textsuperscript{134}In. Vertical dotted lines indicate energies of previously known transitions in~\textsuperscript{134}Sn. (d) The background-subtracted $\gamma$-ray spectrum in~coincidence with the 4041-keV transition in~\textsuperscript{132}Sn observed in the $\beta 2n$ decay of \textsuperscript{134}In. }
\label{fig:134In_coincidences}
\end{figure}

Analysis of $\beta \gamma \gamma$ coincidences reveals three new transitions in~\textsuperscript{134}Sn. The 1666-, 3512- and 3763-keV lines are seen in~spectra of $\gamma$~rays in coincidence 
with previously known transitions in~this nucleus. Figure~\ref{fig:134In_coincidences}a-c displays the~$\gamma$-ray spectra in~coincidence with new transitions assigned to~the daughter nucleus produced in~the $\beta \gamma$-decay branch of~\textsuperscript{134}In. Two~of~them depopulate neutron-unbound states at~excitation energies exceeding the~$S_{1n}$ of~\textsuperscript{134}Sn, 3631(4)~keV~\cite{10.1088/1674-1137/abddaf}, by~more than 1~MeV.  \par
The $\beta 2n$-decay branch of~\textsuperscript{134}In, leading to~the population of~states in~\textsuperscript{132}Sn, was~observed for~the first time. Transitions depopulating the $2^+$, $3^-$, $4^+$ and~$6^+$ states in~\textsuperscript{132}Sn~\cite{10.1103/PhysRevLett.73.2413}, with~energies of~4041, 4351, 375 and~300~keV, respectively, were~identified (see Fig.~\ref{fig:134In_beta-gated_spectrum}). Coincidence relationships observed for~$\gamma$~rays in~the daughter nucleus produced in~the $\beta 2n$-decay branch of~\textsuperscript{134}In are~shown in~Figure~\ref{fig:134In_coincidences}d. 
\par
Transitions assigned to~the $\beta \gamma$-, $\beta 1n$- and $\beta 2n$-decay branches of \textsuperscript{134}In are summarized in~Table~\ref{tab:134In_transitions}. 
\newcommand\T{\rule{0pt}{2.6ex}}       
\newcommand\B{\rule[-1.2ex]{0pt}{0pt}} 
\begin{table}[h]
\renewcommand{\thefootnote}{\alph{footnote}}
\caption{Energies and relative intensities of the transitions observed in~the \textsuperscript{134}In $\beta$~decay. Total $\gamma$-ray and internal-conversion intensities are normalized to the intensity of the 1561-keV transition, for which the absolute intensity is~deduced to be 10.8(6)\% per $\beta$~decay of \textsuperscript{134}In.\\}
\label{tab:134In_transitions}
\begin{tabular}{c c | c c }
\hline \hline
~~ Decay	~~	&		~~ Daughter ~~		&		~~ Energy ~~	&		 ~~ Relative ~~		\T	\\
branch	&		nucleus			&		(keV)		&		intensity			\B \\ \hline  
$\beta \gamma$		&		\textsuperscript{134}Sn & 173.8(3)		&	~4.9(3)\hypertarget{footnote:a}{\hyperlink{footnote_a}{$^a$}}\T\\
$\beta \gamma$		&		\textsuperscript{134}Sn &	347.4(3)		&	~4.9(3)\hypertarget{footnote:a}{\hyperlink{footnote_a}{$^a$}}\\
$\beta \gamma$		&		\textsuperscript{134}Sn &	725.6(3) 		&	4.9(4)\\
$\beta \gamma$		&		\textsuperscript{134}Sn &	1665.5(3) 	&	0.6(1)\\
$\beta \gamma$		&		\textsuperscript{134}Sn &	3512.3(3) 	&	2.7(3)\\
$\beta \gamma$		&		\textsuperscript{134}Sn & 3763(1)			&	~0.5(1)	\B\\ \hline
$\beta 1n$		&		\textsuperscript{133}Sn & 	854.0(3)		&	10.4(7)	\T\\
$\beta 1n$		&		\textsuperscript{133}Sn &	1561.1(3)		&	100(5)	\\
$\beta 1n$		&		\textsuperscript{133}Sn &	2003.8(3)		&	3.7(3)\\
~$\beta 1n$\hypertarget{footnote:b}{\hyperlink{footnote_b}{$^b$}}		&		\textsuperscript{133}Sn & 	2434.2(3)		&	1.4(2)\\
$\beta 1n$		&		\textsuperscript{133}Sn &	3563(1)		&	0.6(2)\\
$\beta 1n$		&		\textsuperscript{133}Sn &	4110(1)		&	0.7(2)\B\\ \hline
$\beta 2n$		&		\textsuperscript{132}Sn &	299.5(3)		&	0.4(2)\T\\
$\beta 2n$		&		\textsuperscript{132}Sn &	375.0(3)		&	0.48(7)\\
$\beta 2n$		&		\textsuperscript{132}Sn &	4041.0(5)		&	0.9(2)\\
$\beta 2n$		&		\textsuperscript{132}Sn &	4351(1)		&	0.5(1)\B\\\hline
\multicolumn{4}{l}{ ~Unassigned:}		\T \B	\\ \hline
Energy  &	Relative	&	Energy  &	Relative		\T\\
(keV)		&	intensity	&	(keV)		& intensity\B\\ \hline
354.3(3)		&	1.5(2)	&		3599(2)	&	~~0.4(1)\hypertarget{footnote:d}{\hyperlink{footnote_d}{$^d$}} \T\\
1427.4(3)		&	0.7(2)	&		~3816(1)\hypertarget{footnote:c}{\hyperlink{footnote_c}{$^c$}}	&	$<0.4$~	\\
1605.8(3)		&	1.0(2)	&		~3824(1)\hypertarget{footnote:c}{\hyperlink{footnote_c}{$^c$}}	&	$<0.4$~		\\
1976.3(3)		&	0.8(1)	&		~3840(1)\hypertarget{footnote:c}{\hyperlink{footnote_c}{$^c$}}	&	$<0.4$~	\\
~2026(1)\hypertarget{footnote:c}{\hyperlink{footnote_c}{$^c$}}		&	$<0.5$~	&		4283(1)	&	0.5(1)	\\
~3251(1)\hypertarget{footnote:c}{\hyperlink{footnote_c}{$^c$}}		&	$<0.4$ \B\\ \hline \hline
\end{tabular}\\\ \\
\hypertarget{footnote_a}{\hyperlink{footnote:a}{$^a$}}
Relative intensities were corrected for internal conversion assuming $E2$ character: $\alpha$\textsubscript{tot}(174~keV)$=$0.227(4) and $\alpha$\textsubscript{tot}(347~keV)$=$0.0221(4)~\cite{BrIcc}.\\
\hypertarget{footnote_b}{\hyperlink{footnote:b}{$^b$}}
See the discussion section for more details on this assignment.\\
\hypertarget{footnote_c}{\hyperlink{footnote:c}{$^c$}}
The identification is~uncertain due to low statistics.\\
\hypertarget{footnote_d}{\hyperlink{footnote:d}{$^d$}}
Upper limit, this intensity includes a~contribution from $SE$~peak. 
\end{table}
Several additional~transitions were observed with a~time pattern consistent with the \textsuperscript{134}In $\beta$-decay half-life. However, due to the lack of $\beta \gamma \gamma$ and $\gamma \gamma$ coincidence relationships, they could not be placed in~the $\beta$-decay scheme of \textsuperscript{134}In. These transitions are also listed in~Table~\ref{tab:134In_transitions}.  \par
\begin{figure*}
\includegraphics[width=\linewidth]{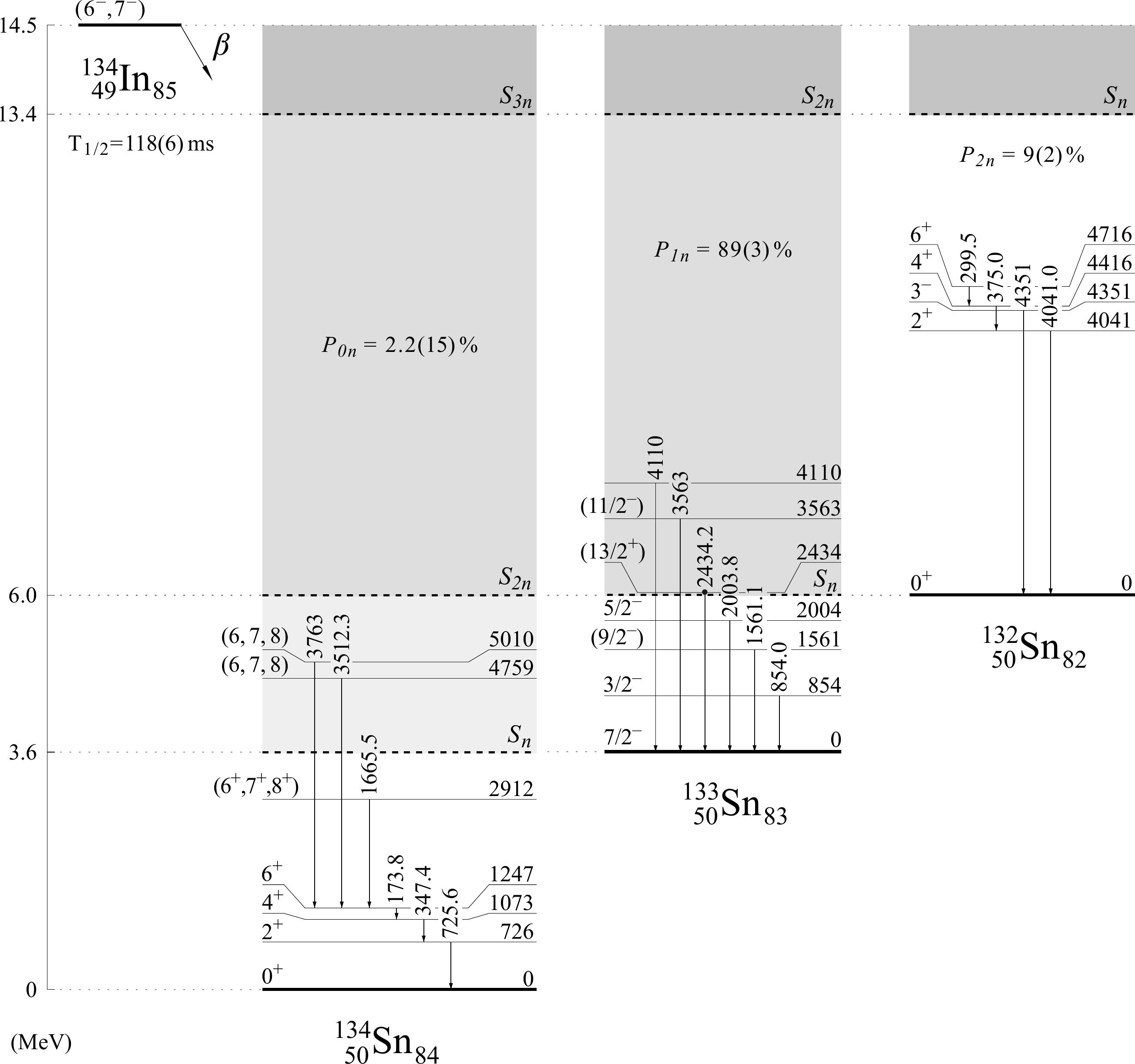}
\caption{Decay scheme of \textsuperscript{134}In established in the present work. Excited states in~the daughter nuclei are labeled with energies (in~keV) given relative to the ground state of each tin isotope. 
The spin-parity assignments for previously known states in~tin isotopes are taken from Refs.~\cite{10.1007/s002180050269, 10.1103/PhysRevLett.77.1020, 10.1103/PhysRevLett.118.202502, 10.1103/PhysRevLett.73.2413}. The~ground-state spin and parity of~\textsuperscript{134}In was~proposed based on~our experimental findings. Shell-model predictions and~systematics discussed in~Sec.~\ref{sec:Discussion_134In-decay} favor the~$7^-$~assignment.
The~left vertical scale (in~MeV) shows the excitation energy and (multi-) neutron separation energies with respect to the \textsuperscript{134}Sn ground-state. The~shaded regions represent energy windows for~population of~(multi-) neutron-unbound states. The $Q_{\beta}$, $S_n$, $S_{2n}$ and $S_{3n}$ values are taken from Ref.~\cite{10.1088/1674-1137/abddaf}.}
\label{fig:134In_decay_scheme}
\end{figure*}

 
The $\beta$-decay scheme of \textsuperscript{134}In established in~the present work is~shown in~Fig.~\ref{fig:134In_decay_scheme}. The previously reported scheme~\cite{10.1103/PhysRevLett.77.1020} is~now complemented by the $\beta \gamma$- and $\beta 2n$-decay branches, with~thirteen new transitions assigned to~this $\beta$~decay. Neutron-unbound states decaying via~$\gamma$~rays were identified in~two daughter nuclei, \textsuperscript{134}Sn and~\textsuperscript{133}Sn. 
It~should be emphasized that presumably only a~partial $\beta$-decay scheme is~established in~this work, since the $\beta$-decay energy of \textsuperscript{134}In is~large ($Q_{\beta}\approx 14.5$\,MeV~\cite{10.1088/1674-1137/abddaf}) and, as~we have presented, the contribution of $\gamma$~ray deexcitation to the decay of neutron-unbound states in~\textsuperscript{134}Sn and~\textsuperscript{133}Sn is~significant. 
\par 
%
Relative intensities of transitions following the \textsuperscript{134}In $\beta$~decay were determined from the $\beta$-gated $\gamma$-ray spectrum. These intensities, normalized to the most intense 1561-keV $\gamma$~ray, agree with those reported in~the previous $\beta$-decay study of \textsuperscript{134}In~\cite{10.1103/PhysRevLett.77.1020, 10.1023/A:1012661816716}. 
%
%
%
For the $\gamma$~rays involved in the 174-347-726~keV cascade decaying from the $6^+$~isomeric state in~\textsuperscript{134}Sn, a~correction to the intensity extracted from the $\beta$-gated $\gamma$-ray spectrum due to~an isomer half-life of~81.7(12)~ns (see Sec.~\ref{sec:Lifetime measurements}) was~included. 
%
The transition intensities determined from the $\beta$-gated $\gamma$-ray spectrum as peak areas corrected for efficiency and internal conversion were found to be equal for the 174-, 347- and 726-keV transitions, suggesting that the $2^+$ and $4^+$ states in~\textsuperscript{134}Sn are not fed directly in~the $\beta$ decay of \textsuperscript{134}In within the intensity uncertainties. 
This~is~further confirmed by the analysis of the $\gamma$-ray spectrum in~coincidence with the 347-keV transition, where the ratio of~transition intensities for~the 174- and~726-keV lines was deduced to be~1.0(1). These observations points to the lack of direct $\beta$-decay feeding to the $2^+$ and $4^+$ states in~\textsuperscript{134}Sn and consequently provides an argument for the high spin value of the ground state of the parent nucleus, which can be $6^-$ or  $7^-$.
 \par 
The probabilities of  $\beta 1n$ and $\beta 2n$ emission from \textsuperscript{134}In were determined from the ratio of~daughter nuclei produced in~a~given $\beta$-decay branch to the total number of~daughter nuclei, using $\gamma$~rays emitted in~their decays. The~following transitions and~their absolute intensities were used: 
872~keV in~\textsuperscript{134}Sb from the \textsuperscript{134}Sn $\beta$~decay with 6(3)\%~\cite{10.1103/PhysRevC.41.R1890}, 341~keV in~\textsuperscript{132}Sb from the \textsuperscript{132}Sn $\beta$~decay with 48.8(12)\%~\cite{10.1016/j.nds.2005.03.001, 10.1103/PhysRevC.39.1963}, and 962~keV in~\textsuperscript{133}Sb from the \textsuperscript{133}Sn $\beta$~decay with 12(2)\%~\cite{10.1007/BF01879878}. For~the latter, both the $\beta$ decay of \textsuperscript{133}Sn and the $\beta n$ decay of~\textsuperscript{134}Sn contribute to~the intensity. For the $\beta n$-decay branch of~\textsuperscript{134}Sn we~use the 1.4\% feeding of~the 962-keV state in~\textsuperscript{133}Sb reported in~Ref.~\cite{10.1103/PhysRevC.41.R1890}. The~$\gamma$-ray intensities obtained from the singles  $\gamma$-ray spectrum were used to~derive the probabilities. Corrections to~the recorded activity of~daughter nuclei due to~tape movement were included based on~the reconstructed average supercycle structure. In~this way we~obtained branching ratios for the $\beta$~decay of~\textsuperscript{134}In: $P_{0n}=2.2(15)\%$, $P_{1n}=89(3)\%$ and $P_{2n}=9(2)\%$. The~$P_{1n}$ value obtained in~our estimate is~larger than the $\beta n$-decay branching ratio evaluated from the previous $\beta$-decay study of \textsuperscript{134}In, $P_{n}\approx 65\%$~\cite{10.1103/PhysRevLett.77.1020, 10.1016/j.nds.2020.09.001, 10.1016/j.nds.2004.11.001}. \par
%
%
\subsection{\protect\boldmath$\beta$ decay of \textsuperscript{135}In}
\begin{figure}
\includegraphics[width=\linewidth]{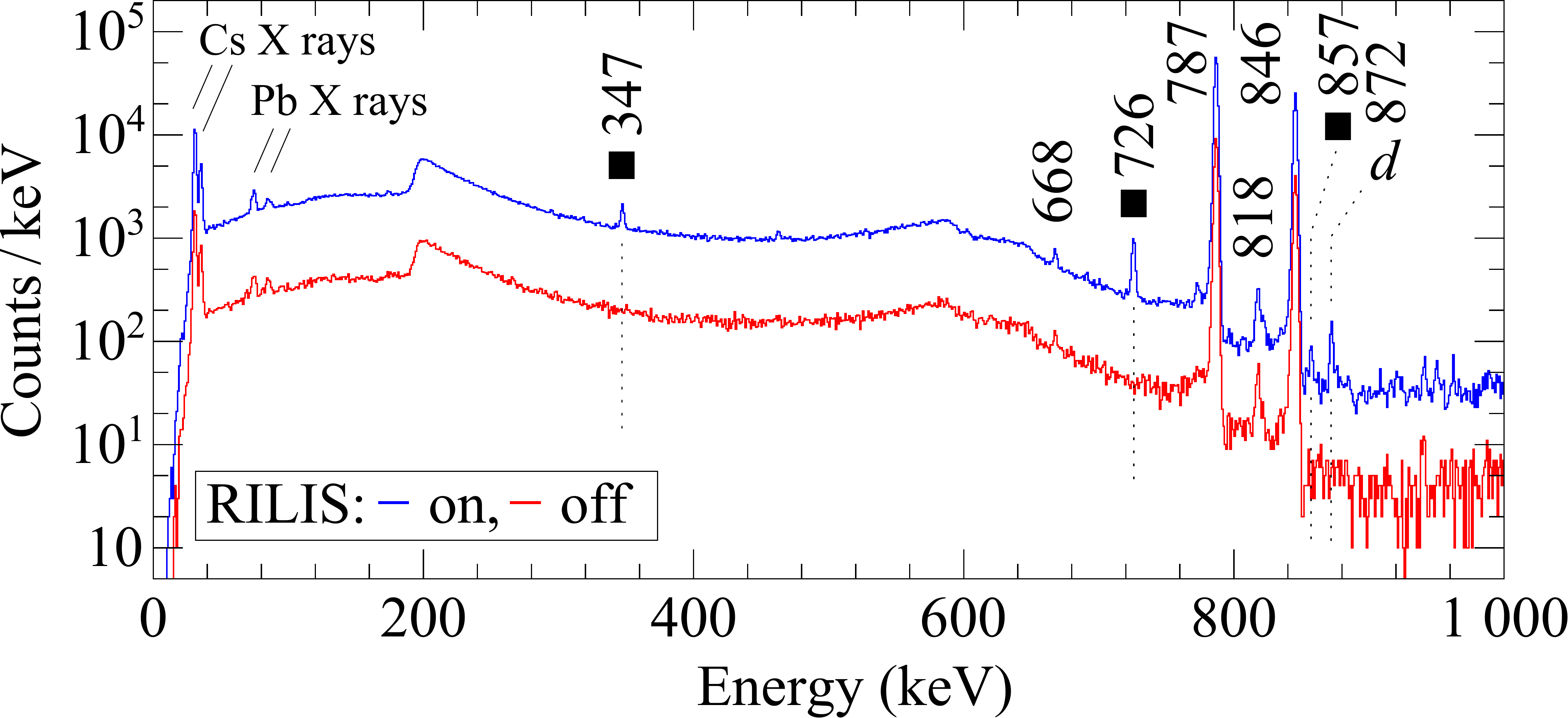}\\
\caption{(Color online) The $\beta$-gated $\gamma$-ray spectrum obtained at $A=135$ when RILIS was applied to ionize indium (upper blue curve) and when one of the RILIS lasers was blocked (lower red curve). Some of the most prominent transitions are labeled with their energies (in~keV). Peaks present in~both spectra originate from the contaminants, while those appearing only in~the RILIS-on mode can be attributed to the $\beta$~decay of \textsuperscript{135}In (square) or its daughter nucleus (marked with  ``$d$"). }
\label{fig:TE-TpClov-beta-135In_0-1MeV_laser_on_off}
~\vspace*{2mm}\\
\includegraphics[width=\linewidth]{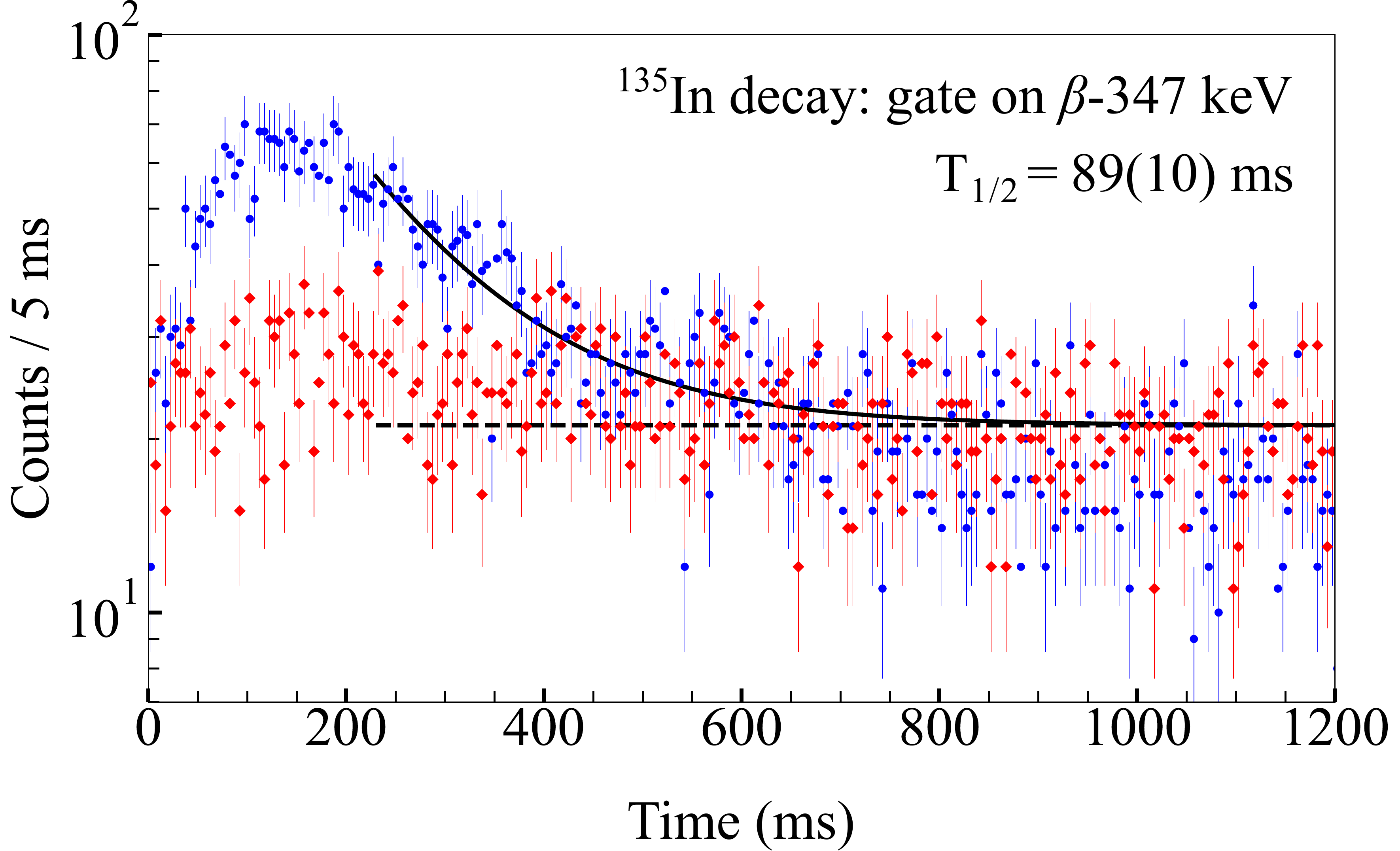}
\caption{(Color online) Time distribution relative to the proton pulse of the 347-keV transition (blue circles) and the background area (red diamonds) observed in~coincidence with $\beta$~particles at $A=135$ in~the laser-on mode. A~function composed of an exponential decay and a~constant background was fitted (solid line) in~the 230-1200~ms time range. The curve corresponding to the background component is~also presented (dashed line). A~Bayesian approach was applied in the statistical analysis of the data~\cite{10.1103/PhysRevC.98.064326}. }
\label{fig:135In_set4_draft}
\end{figure}

Spectra acquired at $A=135$ are dominated by the decay of the surface-ionized \textsuperscript{135}Cs. Figure~\ref{fig:TE-TpClov-beta-135In_0-1MeV_laser_on_off} shows a~comparison of the $\beta$-gated $\gamma$-ray spectra measured in~laser-on and laser-off modes. 
Despite strong isobaric contamination of the RILIS-ionized beam, we~were able to~identify for the first time transitions following the~\textsuperscript{135}In $\beta$~decay. The~two most intense lines seen only in~the spectrum collected when RILIS was used to~ionize indium, at~347 and~726~keV, correspond to~known $\gamma$~rays in~\textsuperscript{134}Sn. 
The $\beta$-decay half-life of~\textsuperscript{135}In was determined from the time distributions of~the 347- and 726-keV transitions which yielded $T_{1/2}=89(10)$~ms and ~$90(9)$~ms, respectively. The~decay curve of~the 347-keV $\gamma$~ray is~shown in~Fig.~\ref{fig:135In_set4_draft}. The~weighted average of 89(7)~ms is~in agreement with the half-life previously determined at~ISOLDE by~measuring the $\beta$-delayed neutrons, 92(10)~ms~\cite{10.1007/s10050-002-8756-7}, and~slightly lower than the half-life of~103(5)~ms measured at~RIKEN~\cite{10.1103/PhysRevLett.114.192501}. 
Based on the systematics of the lighter odd-\emph{A} indium isotopes, a~$\beta$-decaying isomer in~\textsuperscript{135}In is~expected to exist, with a~half-life similar to the ground state~\cite{10.1016/j.nds.2015.08.001}. However, no evidence for its presence was found in~this work. \par
\begin{figure*}\centering
\includegraphics[width=\textwidth]{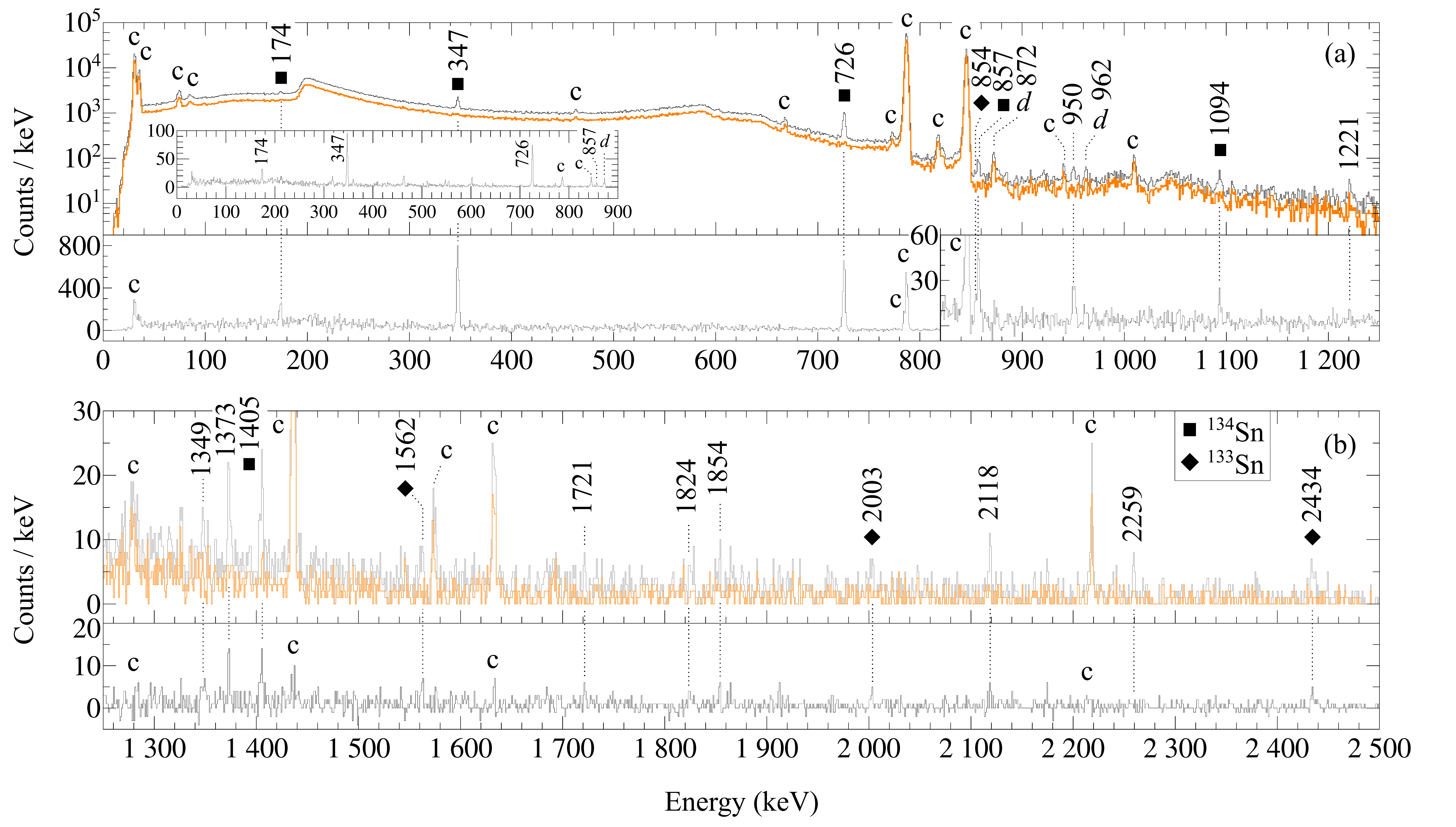}\\
\caption{(Color online) The $\beta$-gated $\gamma$-ray spectra obtained at~$A=135$ in~the laser-on mode in~which different conditions on~time with respect to the proton pulse were applied. 
[Top panels of (a) and (b):] The orange (gray) curve shows the spectrum gated at~times later than 600~ms relative to the proton pulse, while the black curve shows the spectrum without any condition imposed on the time of the event with respect to the proton pulse. 
The inset in (a) shows a~portion of the spectrum with an increased energy threshold for~$\beta$ particles.
[Bottom panels of (a) and (b):] The $\beta$-gated $\gamma$-ray spectrum recorded in~the first 400~ms relative to the~proton pulse from~which long-lived background was subtracted. 
Transitions assigned to the $\beta 1n$- and $\beta 2n$-decay branches of \textsuperscript{135}In are marked with squares and diamonds, respectively.
Peaks that can be attributed to $\gamma$~rays following \textsuperscript{135}In $\beta$~decay are indicated by energy only, while those assigned to activities of the daughter or contaminant nuclei are marked with ``\textit{d}" and ``c", respectively. 
}
\label{fig:135In_spectra_comparison}
\end{figure*}
\begin{figure*}
\includegraphics[width=\linewidth]{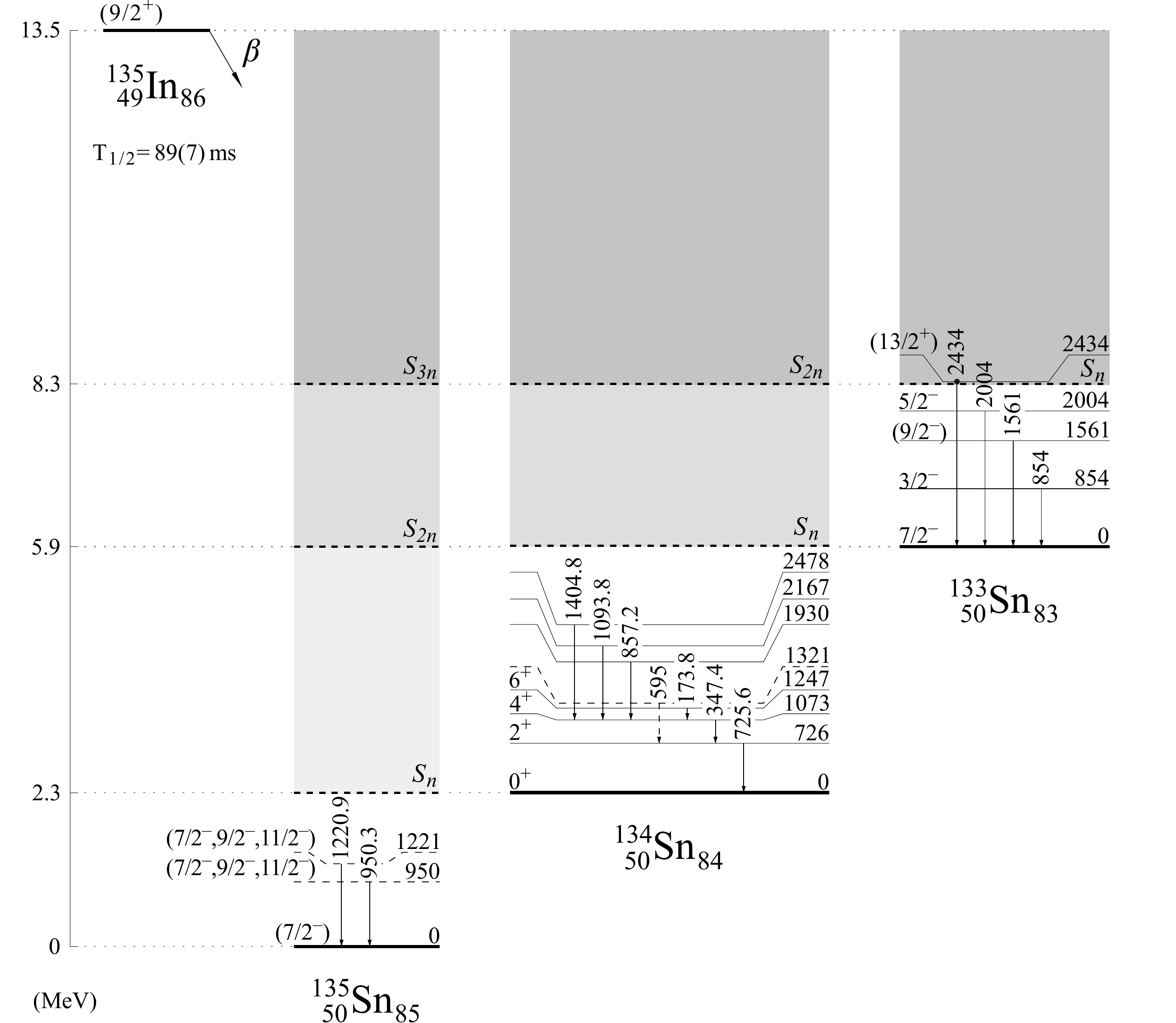}
\caption{Decay scheme of~\textsuperscript{135}In established in~this work. 
Excited states in~daughter nuclei are~labeled with energies (in~keV) given relative to~the ground state of~each tin isotope.
Levels tentatively proposed in~\textsuperscript{135}Sn and~\textsuperscript{134}Sn are~indicated with dashed lines.
The spin-parity assignments for previously known states in~\textsuperscript{134}Sn and \textsuperscript{133}Sn are~taken from Ref.~\cite{10.1007/s002180050269, 10.1103/PhysRevLett.77.1020}. The~ground-state spin and parity of~\textsuperscript{135}Sn and~\textsuperscript{135}In are~based on systematics~\cite{10.1088/1674-1137/abddae}.
The~left vertical scale (in~MeV) shows the~excitation energy and (multi-) neutron separation energies with respect to~the \textsuperscript{135}Sn ground~state. The~shaded regions represent energy windows for~population of~neutron-unbound states. The~$Q_{\beta}$, $S_n$, $S_{2n}$ and~$S_{3n}$ values are taken from~Refs.~\cite{10.1088/1674-1137/abddaf}.
}
\label{fig:135In_decay-scheme}
\end{figure*}

%
%
%
Suppression of the background observed at $A=135$ became crucial for the identification of other transitions following the \textsuperscript{135}In $\beta$~decay. Two~approaches were used independently in~our analysis to~reduce contaminants. 
One~strategy was~to apply a~gate on the first few hundred milliseconds after the proton pulse and subtract events recorded at delayed intervals, leading to a~substantial decrease in~contamination from 53(2)-min \textsuperscript{135$m$}Cs~\cite{10.1088/1674-1137/abddae}.
The second approach was~to study $\gamma$~rays observed in~coincidence with the highest-energy deposit in~the plastic detector, in~order to preferentially select \textsuperscript{135}In $\beta$~decay.
Figure~\ref{fig:135In_spectra_comparison} shows the $\gamma$-ray spectra built using two different $\beta$-gating conditions. By comparing these spectra, transitions following the \textsuperscript{135}In $\beta$~decay were established. Their energies and relative intensities, which were determined from the $\beta$-gated $\gamma$~ray spectrum, are listed in~Table~\ref{tab:135In_transitions}. Figure~\ref{fig:135In_decay-scheme} shows the $\beta$-decay scheme of \textsuperscript{135}In established in~this work. 
\par
\begin{table}
\renewcommand{\thefootnote}{\alph{footnote}}
\caption{Energies and relative intensities of transitions observed in~the \textsuperscript{135}In $\beta$~decay. Total $\gamma$-ray and internal-conversion intensities are normalized to the intensity of the 726-keV transition. \\}
\label{tab:135In_transitions}
\begin{tabular}{c c | c c }
\hline \hline
~~ Decay	~~	&		~~ Daughter ~~		&		~~ Energy ~~	&		 ~~ Relative ~~		\T	\\
branch	&		nucleus			&		(keV)		&		intensity			\B \\ \hline  
$\beta \gamma$\hypertarget{footnote:aa}{\hyperlink{footnote_aa}{$^a$}}
						&		\textsuperscript{135}Sn		&		950.3(3)			&		7(1) 				\T\\
$\beta \gamma$\hypertarget{footnote:aa}{\hyperlink{footnote_aa}{$^a$}}
						&		\textsuperscript{135}Sn		&		1220.9(3)			&		4.0(9) 			\B\\ \hline
$\beta 1n $		&		\textsuperscript{134}Sn		&		 173.8(3)			&		~25(5)\hypertarget{footnote:aa}{\hyperlink{footnote_aa}{$^b$}}			\T\\
$\beta 1n$		&		\textsuperscript{134}Sn		&		347.4(3)			&		~74(5)\hypertarget{footnote:bb}{\hyperlink{footnote_bb}{$^b$}}		\\
~$\beta 1n$\hypertarget{footnote:aa}{\hyperlink{footnote_aa}{$^a$}}		
						&		\textsuperscript{134}Sn		&		~595(1)\hypertarget{footnote:cc}{\hyperlink{footnote_cc}{$^c$}}				&		~11(5)\hypertarget{footnote:dd}{\hyperlink{footnote_dd}{$^d$}}			\\
$\beta 1n $		&		\textsuperscript{134}Sn		&		725.6(3)			&		100(6) 		\\
$\beta 1n $		&		\textsuperscript{134}Sn		&		857.2(3)			&		7(1) 			 \\
$\beta 1n $		&		\textsuperscript{134}Sn		&		1093.8(6)			&		6(1) 			\\
$\beta 1n $		&		\textsuperscript{134}Sn		&		1404.8(6)			&		3.9(8)		\B\\ \hline
$\beta 2n $		&		\textsuperscript{133}Sn		&		854.0(8)			&		~1.6(9)	\T\\
$\beta 2n $		&		\textsuperscript{133}Sn		&		1562.4(8)			&		2.0(6)					\\
$\beta 2n $		&		\textsuperscript{133}Sn		&		2003.3(8)			&		1.8(6) 						\\
~$\beta 2n $\hypertarget{footnote:ee}{\hyperlink{footnote_ee}{$^e$}}	&		\textsuperscript{133}Sn		&		2434.2(7)			&		~2.6(7) 					 \B\\ \hline 
\multicolumn{4}{l}{ ~Unassigned:}		\T \B	\\ \hline
Energy  &	Relative	&	Energy  &	Relative		\T\\
(keV)		&	intensity	&	(keV)		& intensity\B\\ \hline
1349.2(9)			&		2.4(7) 			&		1853.1(8)			&		~2.2(7) 		\T	\\
1372.9(3)			&		3.3(8) 			&		2118.3(6)			&		2.5(7) 				\\
1720.9(8)			&		1.2(5)			& 		2259.3(8)			&		1.8(6)				\\
1824.0(8)			&		2.0(7)			& 		2516.1(8)			&		~1.5(5)		\B	\\ \hline \hline
\end{tabular}\\\ \\
\hypertarget{footnote_aa}{\hyperlink{footnote:aa}{$^a$}} Tentatively assigned to this $\beta$-decay branch of~\textsuperscript{135}In.\\
\hypertarget{footnote_bb}{\hyperlink{footnote:bb}{$^b$}}
Relative intensities were corrected for internal conversion assuming $E2$ character: $\alpha$\textsubscript{tot}(174~keV)$=$0.227(4) and $\alpha$\textsubscript{tot}(347~keV)$=$0.0221(4)~\cite{BrIcc}.\\
\hypertarget{footnote_cc}{\hyperlink{footnote:cc}{$^c$}}
Transition observed only in~$\beta \gamma \gamma$ coincidence.\\
\hypertarget{footnote_dd}{\hyperlink{footnote:dd}{$^d$}}
Intensity obtained from coincidences.\\
\hypertarget{footnote_ee}{\hyperlink{footnote:ee}{$^e$}}
See the discussion section for more details on this assignment.\\
\end{table}
The most intense transitions observed in~the \textsuperscript{135}In $\beta$~decay belong to~\textsuperscript{134}Sn. Three lines that can be~attributed to~the previously known $\gamma$~rays in~\textsuperscript{133}Sn were also identified. The~2434-keV transition, which~was seen in~the \textsuperscript{134}In $\beta$~decay, was~also observed in~the \textsuperscript{135}In $\beta$~decay and is~a~plausible candidate for a~new transition in~\textsuperscript{133}Sn. As~for the possible $\beta 3n$-decay branch of~\textsuperscript{135}In, a~slight excess of~counts over background appears in~the $\gamma$-ray spectrum around 4041~keV, corresponding to the energy of the first-excited state in~\textsuperscript{132}Sn~\cite{10.1103/PhysRevLett.73.2413, 10.1103/PhysRevC.102.014328}. The low statistics does not allow it~to be firmly established whether the $\beta 3n$-decay branch has been observed in~this work for \textsuperscript{135}In. \par
\begin{figure*}\centering
\includegraphics[width=\linewidth]{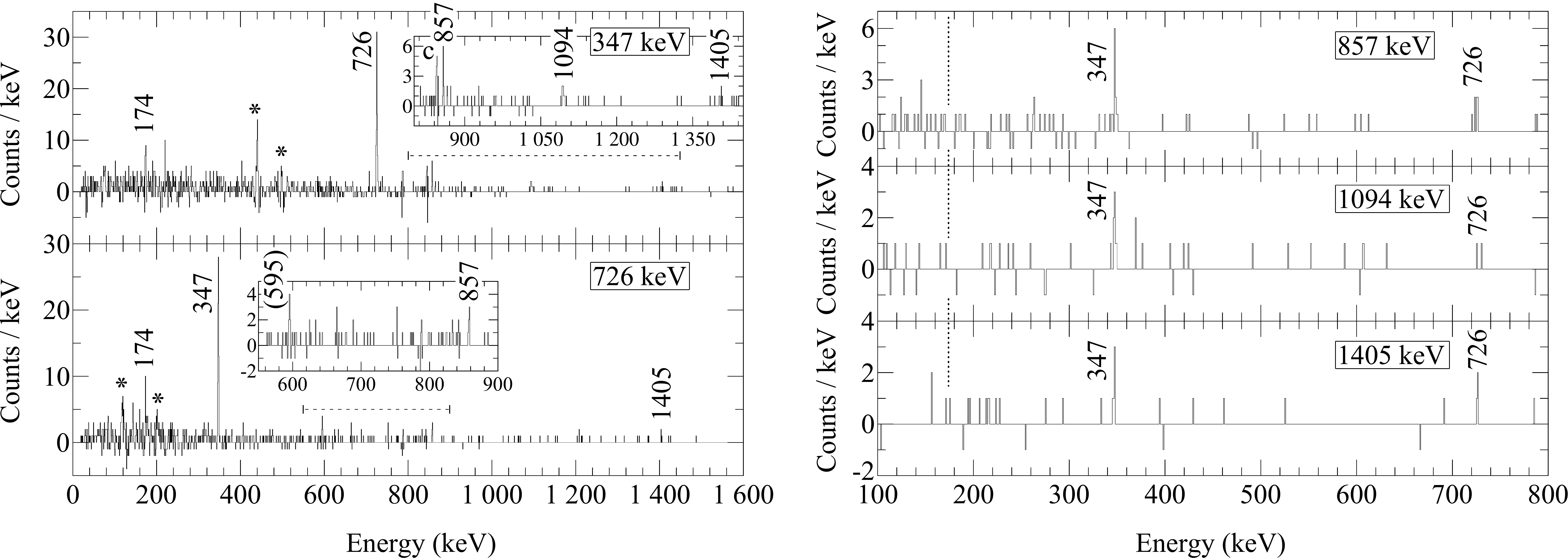}
\caption{(Left panel) Background-subtracted, $\beta$-gated $\gamma$-ray spectra in~coincidence with the 347- and  726-keV transitions that depopulate previously known levels in~\textsuperscript{134}Sn following the $\beta 1n$~decay of \textsuperscript{135}In. Peaks that can be attributed to new transitions in~\textsuperscript{134}Sn are labeled with their energies in~keV. Tentative assignments are given within brackets. Asterisks indicate artifacts due to the background subtraction procedure. The insets display expanded regions whose ranges are indicated with dashed lines. (Right panel) Background-subtracted, $\beta$-gated $\gamma$-ray spectra in~coincidence with newly identified transitions in~\textsuperscript{134}Sn observed in the $\beta$~decay of \textsuperscript{135}In. }
\label{fig:135In_all_coincidences}
\end{figure*}

%
%
Using $\beta \gamma \gamma$ coincidence data, new transitions were identified in~\textsuperscript{134}Sn. Figure~\ref{fig:135In_all_coincidences} displays the $\beta$-gated $\gamma$-ray spectra in~coincidence with the 347- and 726-keV transitions that reveal three new $\gamma$~rays in~\textsuperscript{134}Sn with energies of~857, 1094 and 1405~keV. These transitions were placed in~the level scheme of~\textsuperscript{134}Sn as~depopulating levels at~excitation energies of~1930, 2167 and~2478~keV, respectively (see Fig.~\ref{fig:135In_decay-scheme}). Tentative assignment to~\textsuperscript{134}Sn was made for~the 595-keV transition, which~was found in~coincidence with that at 726~keV, but~was not observed in~the $\gamma$-ray spectra sorted with two different $\beta$-gating conditions (see Fig.~\ref{fig:135In_spectra_comparison}). \par
Several new lines, which~were not observed in~the $\beta$~decays of~the lighter indium isotopes, were seen in~the \textsuperscript{135}In $\beta$~decay. They are~listed in~Table~\ref{tab:135In_transitions}. Based on~the available experimental information on~daughter nuclei produced in~the $\beta 1n$- and~$\beta 2n$-decay branches of~\textsuperscript{135}In, at~least two of~them can be~considered as~transitions in~\textsuperscript{135}Sn. For~\textsuperscript{134}Sn, identification of~new levels below the excitation energy of~the $6^+$ state (at 1247~keV) is~unlikely~\cite{10.1007/s002180050269, 10.1007/PL00013594, 10.1103/PhysRevC.86.054319}. For~\textsuperscript{133}Sn, new~levels below 2004~keV are~also not expected~\cite{10.1103/PhysRevLett.77.1020, 10.1023/A:1012661816716, 10.1103/PhysRevC.84.034601, 10.1103/PhysRevLett.112.172701, 10.1103/PhysRevLett.118.202502, 10.5506/APhysPolB.49.523, 10.1103/PhysRevC.99.024304}. Therefore, the 950- and 1221-keV lines, being the most intense ones in~the considered energy range and for~which no~coincident $\gamma$~rays were observed, were attributed to deexcitations in~\textsuperscript{135}Sn. 
Due to the~higher excitation energies of other transitions as well as the lack of $\beta \gamma \gamma$ and $\gamma \gamma$ coincidences for them, it~was not possible to attribute them to \textsuperscript{135}Sn or \textsuperscript{134}Sn. 
\par
Due~to~the overwhelming~long-lived background in~the singles and $\beta$-gated $\gamma$-ray spectra, evaluation of~the intensities of~$\gamma$~rays following $\beta$~decays of~tin isotopes was not~possible. Thus, absolute intensities of~transitions assigned to~the \textsuperscript{135}In $\beta$~decay could not be~determined. Based on~relative transition intensities, it~can be concluded that the \textsuperscript{135}In $\beta$~decay is~dominated by~the $\beta 1n$~emission.
\par \medskip
\subsection{Lifetime measurements for \textsuperscript{134}Sn}
\label{sec:Lifetime measurements}
For the three lowest excited states in~\textsuperscript{134}Sn, it~was possible to measure their lifetimes using data from both the \textsuperscript{134}In and \textsuperscript{135}In $\beta$~decays.
The fast-timing analyses of~these two $\beta$~decays have their own limitations. In~the case of~\textsuperscript{134}In, acquiring high statistics for transitions in~\textsuperscript{134}Sn was~limited by~the large $P_{1n}=89(3)\%$  and~$P_{2n}=9(2)\%$ values for~the parent nucleus. For~this reason, it~was beneficial to~include in~the lifetime analysis the data collected for~\textsuperscript{135}In, despite the beam contamination problems and over an~order of~magnitude fewer implanted ions of~\textsuperscript{135}In than~\textsuperscript{134}In. The~statistics obtained in these two $\beta$~decays precluded the use of triple coincidences $\beta \gamma \gamma$(t). Nevertheless, by investigating the time response of the background and introducing relevant corrections~\cite{10.1088/1361-6471/aa8217, 10.1103/PhysRevC.102.014328}, half-lives were determined using double-coincidence events~\cite{JB-thesis}. \par
\begin{figure}[t]
\includegraphics[width=\linewidth]{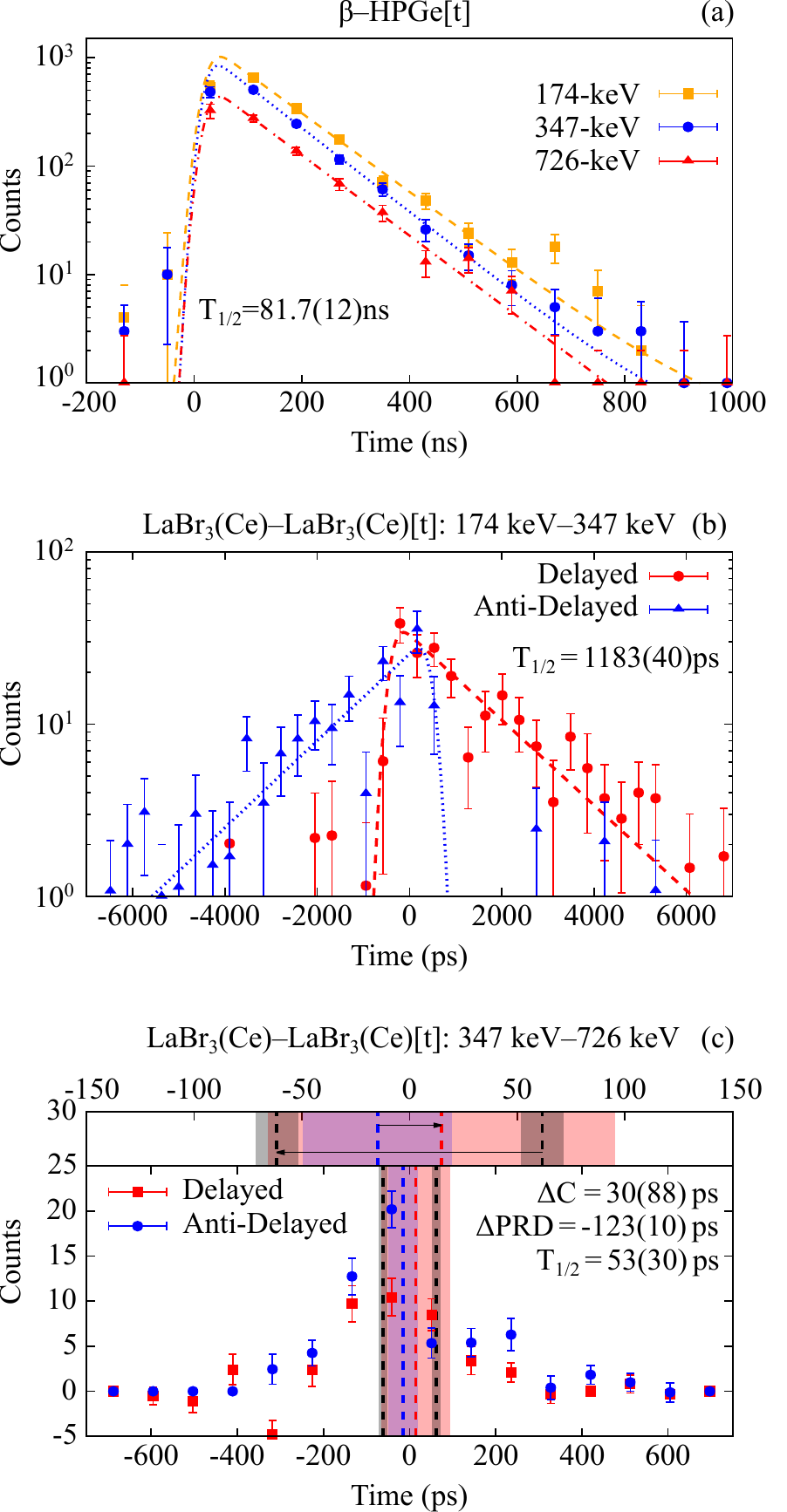}
\caption{Time spectra used to measure the lifetimes of the (a)~$6^+$ 1247-keV, (b)~$4^+$ 1073-keV  and (c)~$2^+$ 726-keV states in~\textsuperscript{134}Sn. In~panels (a) and (b), the half-life is~derived from fit of the slopes, while in~panel (c), it is extracted from the centroid-shift measured between the delayed and antidelayed time spectra ($\Delta C$), which is~caused by the lifetime of the level and the shift in~the Prompt Response Distribution curves ($\Delta PRD$)~\cite{10.1088/1361-6471/aa8217, 10.1103/PhysRevC.102.014328, 10.1088/1361-6471/aa6015}.
See the text for details. 
}
\label{fig:Lifetimes_6-4-2_134Sn_from_134In_JB}
\end{figure}

The half-life of the $6^+$ 1247-keV state in~\textsuperscript{134}Sn was previously reported as $80(15)$\,ns~\cite{10.1007/s002180050269} and, more recently, as~$86^{+8}_{-7}$\,ns~ \cite{10.1103/PhysRevC.86.054319}. Such a~long half-life can be measured using the timing information from the HPGe detectors. Figure~\ref{fig:Lifetimes_6-4-2_134Sn_from_134In_JB}a shows the $\beta -\gamma _{HPGe}(t)$ time distributions gated on the 174-, 347- and 726-keV transitions forming the $6^+\rightarrow 4^+ \rightarrow 2^+\rightarrow 0^+_{g.s}$ cascade from which the half-life of the $6^+$ level was determined to be $81.7(12)$~ns. This value is~in~agreement with those previously reported, but~has a~significantly-improved precision. \par
Determination of the lifetime for the $4^+$ 1073-keV level in~\textsuperscript{134}Sn requires the use of fast $\gamma$~ray detectors. The~$\gamma _{LaBr_{3}(Ce)}-\gamma _{LaBr_{3}(Ce)}$(t) coincidences observed in~the \textsuperscript{134}In $\beta$~decay between two scintillation detectors were used to obtain the time difference between the 174-keV transition feeding the $4^+$ state at 1073~keV and the 347-keV transition depopulating it. Figure~\ref{fig:Lifetimes_6-4-2_134Sn_from_134In_JB}b displays the resulting delayed and antidelayed time distributions. In~the \textsuperscript{135}In $\beta$~decay,  the $6^+$ isomeric state in~\textsuperscript{134}Sn is~weakly populated and the same approach was~not possible. In this case, the lifetime was derived by analyzing the  $\beta -\gamma_{LaBr\textsubscript{3}(Ce)}(t)$ time distributions gated on the 347- and 726-keV $\gamma$~rays depopulating the $4^+$ and $2^+$ states in~\textsuperscript{134}Sn, respectively.
%
By~combining the results from both $\beta$~decays, the half-life of the $4^+$ state in~\textsuperscript{134}Sn was measured for the first time and~determined to be~1183(40)~ps. \par
For~the $2^+$ 726-keV state in~\textsuperscript{134}Sn, the lifetime has not been directly measured to date. A~half-life of 49(7)\,ps was deduced from $B(E2; 0^+\rightarrow 2^+)=0.029(5)$~e$^2$b$^2$ obtained in~a~Coulomb excitation measurement~\cite{10.1016/j.nuclphysa.2004.09.143}. 
To~extract the half-life of the $2^+$ state, the $\gamma_{LaBr\textsubscript{3}(Ce)}-\gamma_{LaBr\textsubscript{3}(Ce)}(t)$ coincidences between the 347- and 726-keV $\gamma$~rays were analyzed. Due to the limited statistics, this was~only feasible using the \textsuperscript{134}In $\beta$-decay data. The~determined centroid positions suffered from low statistics. Figure~\ref{fig:Lifetimes_6-4-2_134Sn_from_134In_JB}c shows the time distributions for~the $347-726$~keV delayed and antidelayed coincidences from which a~half-life of~53(30)~ps was determined for~the $2^+$~state. This~value is~consistent with the one~deduced from the Coulomb excitation measurement~\cite{10.1016/j.nuclphysa.2004.09.143}. 
\par
\section{Discussion}
\subsection{\protect\boldmath$\beta$ decay of \textsuperscript{134}In}
\label{sec:Discussion_134In-decay}
\begin{figure*}
\includegraphics[width=\linewidth]{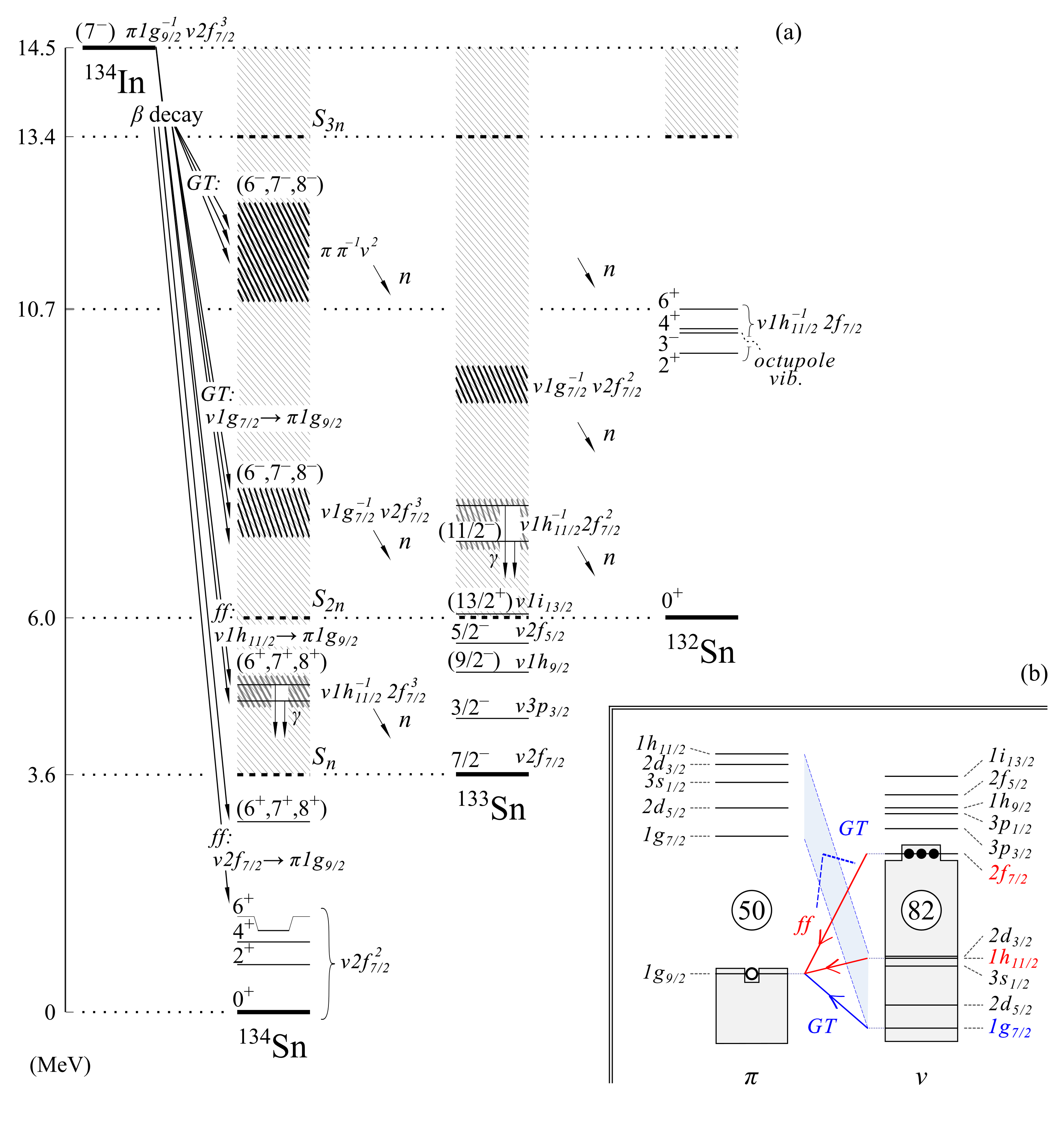}
\caption{(a)~Schematic $\beta$-decay scheme of \textsuperscript{134}In showing Gamow-Teller (GT) and first-forbidden (\emph{ff}) transitions that populate neutron-unbound (gray striped areas) and bound states in~the daughter nuclei. The expected excitation energies of unobserved states having core-excited configurations are indicated by black striped areas. Neutron-unbound states for which decay via~$\gamma$-ray emission was observed are indicated.
 (b) Schematic representation of~proton ($\pi$) and neutron ($\nu$) orbitals relevant for the~$\beta$~decay of \textsuperscript{134}In~\cite{10.1088/0034-4885/70/9/R02}. The~ground-state configuration of~the parent nucleus is~schematically represented by~circles indicating the location of~valence neutrons (full circles) and proton hole (open circle) relative to~the \textsuperscript{132}Sn core.
}
\label{fig:134In-decay-discussion}
\end{figure*}

The ground-state configuration of~\textsuperscript{134}In, with~$Z=49$ and~$N=85$, is~based on~the coupling of the proton hole in~the $\pi 1 g_{9/2}$ orbital and~three neutrons in~the $\nu 2 f_{7/2}$~orbital (see Fig.~\ref{fig:134In-decay-discussion}). 
The \textsuperscript{134}In ground-state spin and~parity have not been determined experimentally~yet. However, in~the previous $\beta$-decay study of \textsuperscript{134}In, it~was possible to restrict the expected spin-parity values to a~range from $4^-$ to~$7^-$, with $7^-$~being favored, based on the observed $\beta$-decay feeding to excited states in~\textsuperscript{133}Sn and on~systematics~\cite{10.1103/PhysRevLett.77.1020}.
The~observation of the~$\beta$-decay feeding to~only one member of~the $\nu 2f_{7/2}^2$ multiplet in~\textsuperscript{134}Sn, the~one with maximum spin value of~$6^+$, is~another argument for a~high ground-state spin of~the parent nucleus, which can be~$6^-$ or~$7^-$.
For~the analogous configuration $\pi 1g_{9/2}^{-1} \,\nu 2 f_{7/2}$ in~\textsuperscript{132}In, the~$7^-$~state is~the lowest-lying member of the multiplet~\cite{10.1103/PhysRevLett.73.2413, 10.1103/PhysRevC.93.041301}. Thus, the~$7^-$ ground state can~also be~expected in~\textsuperscript{134}In. 
This~ground-state spin-parity assignment is~supported by shell-model calculations that reproduce the recently identified $3.5(4)$-$\mu$s isomer in~\textsuperscript{134}In decaying by an~$E2$ transition~\cite{10.1103/PhysRevC.100.0113020}. 
Shell-model calculations with two different interactions consistently predict $7^-$ as~the \textsuperscript{134}In ground state, while the $6^-$ state is expected to lie above the $5^-$ isomer~\cite{10.1103/PhysRevC.100.0113020}. Therefore, we~consider $7^-$ as~the most likely ground-state spin-parity assignment for~\textsuperscript{134}In. \par
The $\beta$~decay of \textsuperscript{134}In is~dominated by the Gamow-Teller (GT) $\nu 1g_{7/2} \rightarrow \pi 1g_{9/2}$ transition~\cite{10.1103/PhysRevC.67.025802}. Since this GT~decay involves deeply-bound neutrons in~the $N=82$ \textsuperscript{132}Sn core, it~populates neutron-unbound states in~the daughter nucleus (see Fig.~\ref{fig:134In-decay-discussion}). These are expected in~\textsuperscript{134}Sn at~excitation energies comparable to~the energy of~the $6^-$~state in~\textsuperscript{132}Sn (7211~keV), arising from the $\nu 1g_{7/2}^{-1} 2f_{7/2}$ configuration, which is~populated in~the GT~decays of~\textsuperscript{132}In~\cite{10.1103/PhysRevLett.73.2413, 10.1103/PhysRevC.102.014328}. 
This implies that the prevalent $\beta$-decay feeding is~located well above the $S_{2n}$ of~\textsuperscript{134}Sn, 6030(4)~keV~\cite{10.1088/1674-1137/abddaf}. As~a~result, the~\textsuperscript{134}In $\beta$~decay proceeds mainly through $\beta n$-emission branches.
The~observed population of the $6^+$ 4716-keV state in~the $\beta 2n$-decay daughter nucleus \textsuperscript{132}Sn indicates that there is~significant $\beta$-decay feeding to neutron-unbound states in~\textsuperscript{134}Sn at~excitation energies exceeding 10~MeV. 
This~$\beta$-decay strength most likely originates from GT transitions involving proton particle-hole excitations across the $Z=50$ shell gap (see Fig.~\ref{fig:134In-decay-discussion}). \par
The obtained $\beta 1n$- and $\beta 2n$-decay branching ratios for~\textsuperscript{134}In allow for verification of the predictions of the models used for calculating $\beta$-delayed particle emission, which are employed in~$r$-process nucleosynthesis modeling. There are presently only two known $\beta 2n$ emitters in~the \textsuperscript{132}Sn region for which $P_{2n}$ have been measured~\cite{10.1016/j.nds.2020.09.001, IAEA-database}: \textsuperscript{136}Sb with $P_{2n}=0.14(3)\%$~\cite{10.1103/PhysRevC.98.034310} and~\textsuperscript{140}Sb with $P_{2n}=7.6(25)\%$~\cite{10.1103/PhysRevC.95.044322}. 
\par
The $P_{1n}$ and $P_{2n}$ probabilities obtained in~this work are compared with theoretical predictions based on quasiparticle random phase approximation (QRPA)~\cite{10.1006/adnd.1997.0746}, relativistic Hartree-Bogoliubov (RHB) model with the proton-neutron relativistic QRPA (RQRPA)~\cite{10.1103/PhysRevC.93.025805} as well as phenomenological effective density model (EDM)~\cite{10.1103/PhysRevC.88.041301} (see Table~\ref{tab:Pn-predictions}). For the QRPA, it~is possible to compare three successively extended models, some of which take into account not only GT transitions but also first-forbidden (\emph{ff}) transitions and competition between all available decay branches of neutron-unbound states. 
\begin{table}
\caption{Comparison of predicted and experimental values of $P_{1n}$ and $P_{2n}$ for \textsuperscript{134}In. 
Results of calculations using three successively improved approaches based on QRPA: QRPA-1~\cite{10.1006/adnd.1997.0746}, QRPA-2~\cite{10.1103/PhysRevC.67.055802} and QRPA+HF~\cite{10.1016/j.adt.2018.03.003, 10.1103/PhysRevC.94.064317} as well as based on RQRPA~\cite{10.1103/PhysRevC.93.025805} and EDM~\cite{10.1103/PhysRevC.88.041301, 10.1103/PhysRevC.90.054306} are presented. Data were taken from~Ref.~\cite{IAEA-database}. 
Predictions of the EDM model after applying the cutoff model (EDM\textsubscript{\textit{cutoff}})~\cite{10.1103/PhysRevC.88.041301, 10.1103/PhysRevC.90.054306} are also presented.\\
 }
\label{tab:Pn-predictions}
\begin{tabular}{c c c }\hline \hline
Method		& 		~\hspace{3mm} $P_{1n} (\%)$~\hspace{3mm}		& ~\hspace{2mm}$P_{2n} (\%)$~\hspace{2mm} \T \B \\ \hline 
\T
QRPA-1 					&		0.60					&	99.4		\\
QRPA-2 					&		 6.5					&	86.7		\\
QRPA+HF 				&		 78					&	15		\\
~~RHB+RQRPA~~ 	&		 18.9					&	46.8		\\
EDM			 				&		64.5					& 2.2		\\
EDM\textsubscript{\textit{cutoff}}	
								&		28			 			&	39\B\\ \hline 
Experiment:			&		89(3)				& ~9(2) \T \B  \\\hline\hline
\end{tabular}
\end{table}
The~inclusion of \emph{ff} transitions in~the QRPA-2 model~\cite{10.1103/PhysRevC.67.055802} leads to an increase in~the $\beta 1n$-decay branching ratio by a~factor of~about~ten with respect to~the previous model, QRPA-1, in~which only GT transitions were considered~\cite{10.1006/adnd.1997.0746}. A~larger contribution of the $\beta 1n$ emission from \textsuperscript{134}In is~predicted by RHB+RQRPA~\cite{10.1103/PhysRevC.93.025805}, which accounts for both GT and \emph{ff} transitions. However, in~the RHB+RQRPA calculations, the total probability of~$\beta n$ emission ($P_{n,tot}$) is~lower ($\approx 66\%$) than in~the two first variants of~the QRPA calculations, where~$P_{n,tot}$ exceeds~90\%. Besides, the~predicted branching ratio of~the $\beta 1n$ decay remains lower than the experimental result. The dominant contribution of the $\beta 1n$ emission is~predicted by the most recent QRPA calculations in~which the statistical Hauser-Feshbach (HF) model~\cite{10.1103/PhysRevC.94.064317} is~incorporated to address competition between $\gamma$-ray, one- and multiple-neutron emission in~the decay of~neutron-unbound states (QRPA+HF)~\cite{10.1016/j.adt.2018.03.003}. The~$P_{1n}$ value predicted by the QRPA+HF model, which also accounts for \emph{ff} transitions, is~the closest to the experimental value among the models considered. In~the case of the $P_{2n}$, the experimental value is~well reproduced only by the EDM. This approach also accounts for the competition between one- and multiple-neutron emission as well as $\gamma$-ray emission above $S_n$. If~the cutoff method is~applied to the EDM, so that the decay of states above $S_{1n}$ ($S_{2n}$) proceeds only via emission of one (two) neutron(s), the calculated probabilities change significantly, $P_{1n}=28\%$ and $P_{2n}=39\%$~\cite{10.1103/PhysRevC.90.054306}. \par
A~comparison of the different $P_{1n,2n}$ calculations shows that the best reproduction of the experimental values for \textsuperscript{134}In is~achieved when \emph{ff} transitions and all possible deexcitation paths of neutron-unbound states are taken into account.
Indeed, the inclusion of competition between the emission of one and multiple neutrons as well as $\gamma$~rays following the \textsuperscript{134}In $\beta$~decay is~relevant, as~in~this work the $\beta1n$-decay branch of \textsuperscript{134}In was~observed to be dominant even though the GT resonance is~located substantially above $S_{2n}$ of \textsuperscript{134}Sn. Moreover, neutron-unbound states decaying via $\gamma$~rays were observed in~the two daughter nuclei, \textsuperscript{134}Sn and~\textsuperscript{133}Sn. In~the one-neutron knockout reaction from \textsuperscript{134}Sn, it~was estimated that around $25\%-35\%$ of the decay of neutron-unbound states populated in~\textsuperscript{133}Sn proceeds via $\gamma$-ray emission~\cite{10.1103/PhysRevLett.118.202502}. The enhanced $\gamma$-ray emission from states above $S_n$ was explained by the small spectroscopic overlap between states involved in~neutron emission. \par
A~similar nuclear structure effect is~expected to play a~role in the $\beta$~decay of~\textsuperscript{134}In, both in~$\beta \gamma$- and $\beta 1n$-decay branches. The~GT~decays of~neutrons from the $N=82$  \textsuperscript{132}Sn core result in~population of~states in~\textsuperscript{134}Sn formed by~couplings of the valence neutrons to~core excitations ($\nu ^{-1} \nu^3$ or $\pi \pi^{-1} \nu^2$, see Fig.~\ref{fig:134In-decay-discussion}). The~wave functions of~the states populated following neutron emission have small spectroscopic overlaps with the low-lying states in~\textsuperscript{133}Sn, having a~single-particle nature~\cite{10.1038/nature09048}. For~this reason, $\gamma$~rays are~able to~compete with neutron emission well above $S_{1n}$. Similar structure effects leading to~hindrance of~neutron emission were identified in~other $\beta n$~emitters~\cite{10.1103/PhysRevLett.117.142701, 10.1103/PhysRevC.95.024320, 10.1016/j.physletb.2017.06.050}. The~QRPA+HF calculations estimate a~minor change, below $3\%$, in~the calculated $\beta n$ emission probability if an~increase of one order of magnitude to the $\gamma$-ray strength function is~assumed~\cite{10.1103/PhysRevC.94.064317}. However, such enhancement of $\gamma$-ray emission would have a~larger effect on the neutron capture rates of neutron-rich nuclei. \par
The population of states below the excitation energy of 7\,MeV in~\textsuperscript{134}Sn is~due to \emph{ff}~$\beta$~decays of \textsuperscript{134}In. One of these, the $\nu 1h_{11/2} \rightarrow\pi 1g_{9/2}$ transition, which involves neutrons from the $N=82$ \textsuperscript{132}Sn core, feeds neutron-unbound states located below the GT resonance (see~Fig.~\ref{fig:134In-decay-discussion}). 
The two new states identified in~\textsuperscript{134}Sn at excitation energies around 5~MeV are most likely members of the $\nu 1h_{11/2}^{-1} 2f^3_{7/2}$ multiplet. This assignment is~supported by shell-model calculations with core excitations, which predict the first state from this multiplet at around 5~MeV (see Fig.~\ref{fig:134Sn_level_scheme_exp_vs_th_comparison})~\cite{10.1103/PhysRevC.84.044324}. An~analogous $(11/2^-)$ state in~\textsuperscript{133}Sn, resulting from the coupling of a~neutron hole in~the $\nu 1h_{11/2}$ orbital and two neutrons in~the $\nu 2f_{7/2}$ orbital, was identified at~3564~keV~\cite{10.1103/PhysRevLett.77.1020, 10.1103/PhysRevLett.118.202502, 10.1103/PhysRevC.99.024304}. The 1.26-MeV neutrons~\cite{10.1103/PhysRevLett.77.1020} and 3564-keV $\gamma$~rays~\cite{10.1103/PhysRevC.99.024304} were assigned to~the decay of~the $(11/2^-)$ state in~\textsuperscript{133}Sn in~$\beta$-decay studies of~\textsuperscript{133}In. The observation of a~3563(1)-keV transition in this work implies that this neutron-unbound $(11/2^-)$ state is~also populated via the $\beta 1n$ decay of \textsuperscript{134}In. A~certain analogy can be noted to~the population pattern observed in the $\beta n$ decay of \textsuperscript{132}In, which proceeds primarily through the high-spin $(11/2^-)$ isomer in~\textsuperscript{131}Sn~\cite{10.1103/PhysRevC.102.014328, 10.1103/PhysRevC.102.024327}. In~the $\beta$~decay of~\textsuperscript{134}In, states with configurations involving neutron hole in~the $\nu 1h_{11/2}$ orbital are populated in~each observed $\beta$-decay branch. These states are neutron-unbound in~both \textsuperscript{134}Sn and~\textsuperscript{133}Sn. However, $\gamma$-ray deexcitation has a~significant contribution to their decay. This~means that states populated following neutron emission, with hole in~the $\nu 1h_{11/2}$ orbital, have little overlap with low-energy states in~the corresponding $\beta n$-decay daughter nuclei, which correspond to~excitations of~valence neutrons in~the $N=82-126$~shell.  
\par
The large $P_{1n}=89(3)\%$ value and the expected $7^-$ ground-state spin and~parity for~\textsuperscript{134}In set favorable conditions to search for the missing $\nu 1i_{13/2}$ s.\,p.~state in~\textsuperscript{133}Sn. 
The high excitation energies of the predicted multiplets in~\textsuperscript{134}Sn involving the $\nu 1i_{13/2}$ orbital are~also advantageous.~The~lowest-lying state arising from the $\nu 2f_{7/2} 1i_{13/2}$ configuration is~expected at an excitation energy of around 4$-$5\,MeV~\cite{10.1103/PhysRevC.76.024313} or 3.2\,MeV~\cite{10.1103/PhysRevC.65.051306}, where negative-parity particle-hole excitations are also expected to appear.
Due to the negative parity of states involving the $\nu 1i_{13/2}$ orbital and the expected high density of such levels in~\textsuperscript{134}Sn~\cite{10.1103/PhysRevC.65.051306}, there is~a~chance that they are~mixed with other neutron-unbound states of negative parity. Such admixtures would increase the overlap of~the wave functions of~states involved in~the $\beta 1n$ decay in~which the $13/2^+$ state in~\textsuperscript{133}Sn can be~populated. 
Since there is~a~wide range of spins, from $3/2^-$ to~$(11/2^-)$, for~the states populated in~\textsuperscript{133}Sn following the \textsuperscript{134}In~$\beta$~decay, the~population of~the $13/2^+$ state does not seem to be~hindered in~terms of~the angular momentum for~neutron emission.\par
The excitation energy of the first $13/2^+$ level in~\textsuperscript{133}Sn was~estimated to~be 2511(80)~keV~\cite{10.1103/PhysRevC.91.027303} or~between 2360 and~2600~keV~\cite{10.1103/PhysRevC.94.034309}. The~2434-keV transition, which is~the only one registered in~the energy range from 2100 to 3500~keV that can be attributed to the $\beta$~decay of~\textsuperscript{134}In (see Fig.~\ref{fig:134In_beta-gated_spectrum}), is~therefore a~natural candidate for a~transition deexciting the $13/2^+$ state in~\textsuperscript{133}Sn. Due to the large difference between the \textsuperscript{134}In and \textsuperscript{134}Sn ground-state spins, direct or indirect feeding of an~excited state in~\textsuperscript{134}Sn that decays to the $0^+$ ground state is~unlikely in~the \textsuperscript{134}In $\beta$~decay.
The 2434-keV transition is~also observed in the \textsuperscript{135}In $\beta$~decay, in~which other states in~\textsuperscript{133}Sn are populated in~the $\beta 2n$-decay branch. 
\par
The decay of the $13/2^+$ state to~the $7/2^-$ ground state in~\textsuperscript{133}Sn can~proceed via an~$E3$ transition with an expected lifetime of around 2\,ns. 
%
In~the analogous nucleus in~the \textsuperscript{208}Pb region with one neutron above the core, \textsuperscript{209}Pb, a~$15/2^-$ level corresponding to the $\nu 1j_{15/2}$ s.\,p. state decays via an~$E3$ transition to the $9/2^+$ ground state ($\nu 2g_{9/2}$) and via an $M2$ transition to the $11/2^+$ excited state ($\nu 1i_{11/2}$)~\cite{10.1016/0370-2693(67)90224-9, 10.1016/j.nds.2015.05.003}. The observed relative intensities of these two transitions are $100(2)$ and $11(1)$, respectively. 
Relying on the similarity of the corresponding excitations in~the \textsuperscript{132}Sn and \textsuperscript{208}Pb regions~\cite{Blomqvist-cds-cern, 10.1103/PhysRevLett.80.5504, 10.1103/PhysRevC.60.024303, 10.1007/BF01879878, 10.1103/PhysRevC.91.027303, 10.1103/PhysRevC.80.021305}, the $E3$ transition is~anticipated to dominate the decay of the $13/2^+$ state in~\textsuperscript{133}Sn.
%
It~is worth mentioning that a~transition with energy of 2434~keV was identified in~\textsuperscript{131}Sn~\cite{10.1103/PhysRevC.70.034312}. However, an~excited state with that energy cannot be populated in~\textsuperscript{131}Sn following the~\textsuperscript{134}In $\beta$ decay due to~an~insufficient $\beta$-decay energy window. 
\par
For the three newly identified states in~\textsuperscript{134}Sn, populated by the \textsuperscript{134}In $\beta$ decay, it~is possible to propose their spins taking into account the observed $\gamma$-ray depopulation pattern and the favored $7^-$ ground-state spin-parity assignment for the parent nucleus. Spin values for~the 2912-, 4759- and~5010-keV levels can be~limited to a~range from 6 to~8, since their decay to~the $6^+$ state at~1247~keV was~observed, while~the $\gamma$-ray decay branch to~the $4^+$ level at~1073~keV was~not identified. For~the state at~2912~keV, a~positive parity can also be~proposed. Due~to the nature of~the low-lying neutron s.\,p.~orbitals in~the $N=82-126$ shell, the~bound states in~\textsuperscript{134}Sn can be~populated solely via~\textit{ff}~decays of~\textsuperscript{134}In (see Fig.~\ref{fig:134In-decay-discussion}). \par
A particular remark should be made about the 354-keV transition, which is~confirmed in this work as following the $\beta$~decay of \textsuperscript{134}In~\cite{10.1103/PhysRevLett.77.1020}. 
Due~to the lack of $\beta \gamma\gamma$ or $\gamma\gamma$ coincidence relations, its~assignment to one of the daughter nuclei is~not possible. A~state decaying directly to the ground state cannot be placed at such a~low excitation energy in~the level scheme of the \textsuperscript{132--134}Sn isotopes. 
In view of the enhanced contribution of electromagnetic transitions above $S_n$ in~\textsuperscript{133}Sn and \textsuperscript{134}Sn, one might consider the possibility that the 354-keV $\gamma$ rays are emitted from a~neutron-unbound state for which the centrifugal barrier hinders neutron emission. Once a~$\gamma$~ray has been emitted with the associated angular-momentum transfer, the level that has been fed could subsequently decay via neutron emission. \par
\subsection{\protect\boldmath$\beta$ decay of  \textsuperscript{135}In}
The $\beta$-decay feeding pattern of the $N=86$ \textsuperscript{135}In is~expected to be similar to that observed in the $\beta$ decay of the $N=84$ \textsuperscript{133}In~\cite{10.1103/PhysRevC.99.024304}. The ground state of \textsuperscript{133}In has a~$\pi  1g_{9/2}^{-1} \, \nu 2f_{7/2}^{^2}$ configuration, while for the ground state of \textsuperscript{135}In, an~additional pair of neutrons occupies the $\nu 2f_{7/2}$ orbital. Based on systematics of the $Z=49$ isotopes~\cite{10.1088/1674-1137/abddae}, a~$9/2^+$ ground-state spin-parity assignment is~expected for both \textsuperscript{133}In and \textsuperscript{135}In. For~the \textsuperscript{133}In nucleus, this~spin value is~supported by~the observed $\beta$-decay feeding to~levels in~\textsuperscript{133}Sn~\cite{10.1103/PhysRevC.99.024304} with well-established spins and~parities~\cite{10.1038/nature09048, 10.1103/PhysRevLett.112.172701}. 
\par
As~discussed in the previous section for \textsuperscript{134}In, the $\beta$~decays of neutron-rich indium isotopes with  $N>82$ are dominated by the GT $\nu 1g_{7/2} \rightarrow \pi 1g_{9/2}$ transition populating states above $S_{1n}$ in~the daughter nuclei. Therefore, the~\textsuperscript{135}In $\beta$~decay is also~dominated by the $\beta n$-decay branches, as~was observed in~this work.
The analogous state attributed to this GT decay was proposed in~\textsuperscript{133}Sn at an~excitation energy of around 6\,MeV~\cite{10.1103/PhysRevC.99.024304}. The lowest-lying states populated via the $\nu 1g_{7/2} \rightarrow \pi 1g_{9/2}$ $\beta$~decay can be expected in~\textsuperscript{135}Sn at comparable energies, being close to the $S_{2n}$ of 5901(4)~keV~\cite{10.1088/1674-1137/abddaf}. Based on the observations from the $\beta$~decay of \textsuperscript{134}In, other GT transitions involving deeply-bound orbitals in~the \textsuperscript{132}Sn core also contribute, which enhances the $\beta 1n$- and $\beta 2n$-decay branches of \textsuperscript{135}In. \par
While~the states populated via~the dominant GT~decays of~\textsuperscript{135}In are~mainly due to particle-hole excitations across the $N=82$~shell gap, levels at~low excitation energies in~\textsuperscript{135}Sn can be~interpreted as~excitations involving neutron orbitals in~the $N=82-126$ shell.
In~analogous $\beta$~decay of the $(9/2^+)$ \textsuperscript{133}In ground state, only two bound states in~\textsuperscript{133}Sn are populated: the $7/2^-$ ($\nu 2f_{7/2}$)  ground state and the $(9/2^-)$ ($\nu 1h_{9/2}$) excited state~\cite{10.1103/PhysRevC.99.024304}. 
Since the structure of the three valence-particles nucleus \textsuperscript{135}Sn is~more complex than the one valence-particle nucleus~\textsuperscript{133}Sn, more bound states can be populated via \emph{ff} transitions in~\textsuperscript{135}Sn than in~\textsuperscript{133}Sn. If~we make an analogy to~the \textsuperscript{133}In~$\beta$~decay~\cite{10.1103/PhysRevC.99.024304}, then~the population of~states arising from the $\nu 2f^3_{7/2}$ and~$\nu 2f_{7/2}^{2} 1h_{9/2}$ configurations in~\textsuperscript{135}Sn is~expected in~the~\textsuperscript{135}In $\beta$~decay.
Taking into account the most probable $(9/2^+)$ ground-state spin of \textsuperscript{135}In, the \emph{ff}-type $\beta$~decay should favor the population of $7/2^-$, $9/2^-$ and~$11/2^-$ states in~\textsuperscript{135}Sn. 
Therefore, the 950- and 1221-keV transitions observed in~the \textsuperscript{135}In $\beta$~decay are assigned as~deexciting states in~\textsuperscript{135}Sn with proposed spin-parity values of~$7/2^-$, $9/2^-$ or $11/2^-$.
\subsection{Comparison with shell-model calculations}
\subsubsection*{\textsuperscript{134}Sn}
\begin{figure*}
\includegraphics[width=\linewidth]{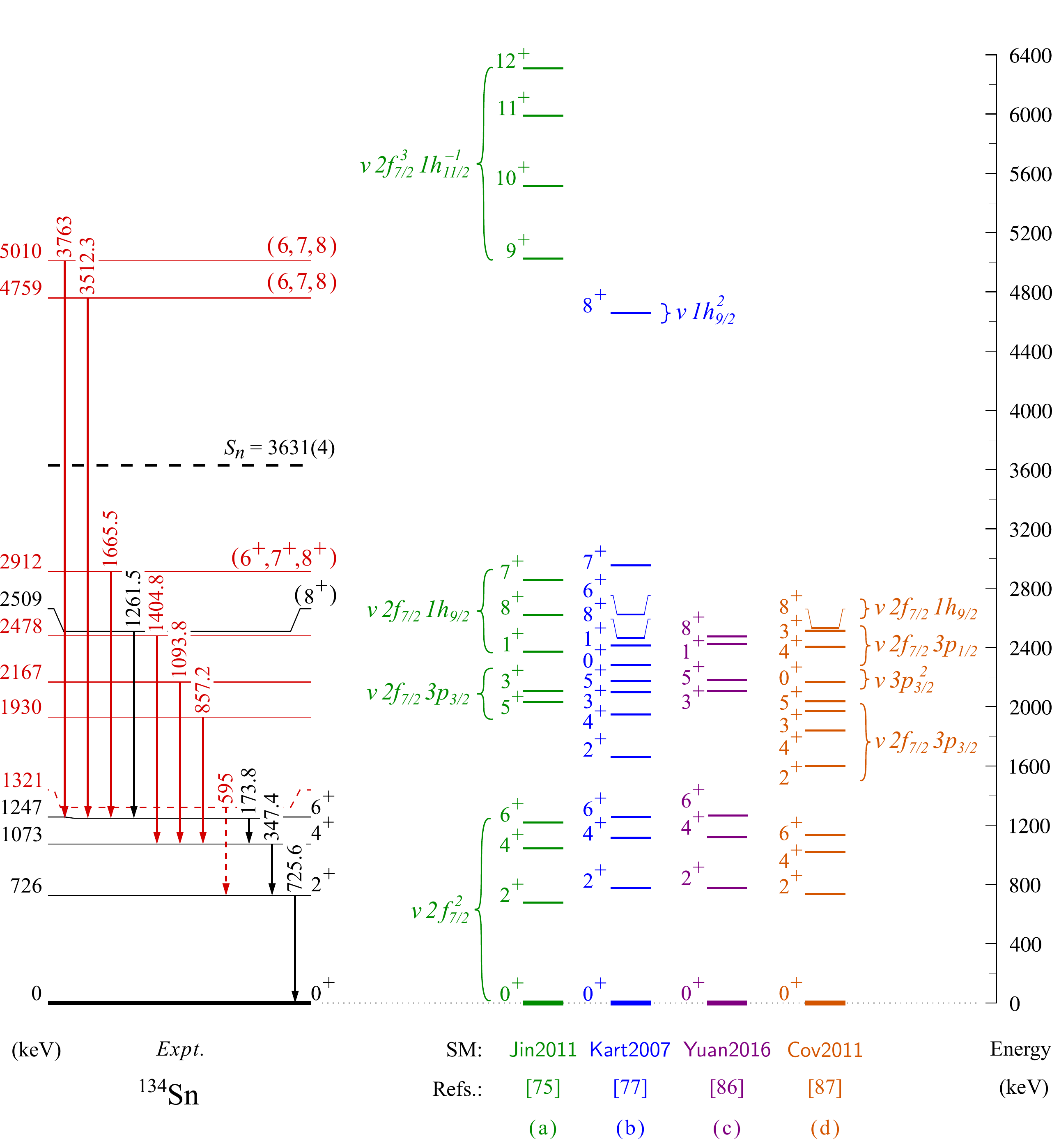}
\caption{Experimental (\textit{Expt.}) level scheme of~\textsuperscript{134}Sn along with the results of~the shell-model calculations (SM) (a)~including neutron-core excitations, \textsf{Jin2011} from Ref.~\cite{10.1103/PhysRevC.84.044324}, as well as employing \textsuperscript{132}Sn as a~closed core: (b) \textsf{Kart2007}~from Ref.~\cite{10.1103/PhysRevC.76.024313}, (c)~\textsf{Yuan2016}~from Ref.~\cite{10.1016/j.physletb.2016.09.030} and (d)~\textsf{Cov2011} from Ref.~\cite{10.1088/1742-6596/267/1/012019}. 
The newly identified states are indicated in~red. The level shown by~the dashed line is~proposed tentatively.
The~$(8^+)$ state at~2509~keV~\cite{10.1007/PL00013594} was not~observed in~this work. 
The~experimental spin-parity assignments for~previously known~states in~\textsuperscript{134}Sn were taken from~Refs.~\cite{10.1007/s002180050269, 10.1007/PL00013594}. 
The $S_n$ value for~\textsuperscript{134}Sn was~taken from Ref.~\cite{10.1088/1674-1137/abddaf}.
}
\label{fig:134Sn_level_scheme_exp_vs_th_comparison}
\end{figure*}

Shell-model predictions for \textsuperscript{134}Sn are compared with the excited states observed in~this nucleus in~Fig.~\ref{fig:134Sn_level_scheme_exp_vs_th_comparison}. 
%
%
The previously reported states in~\textsuperscript{134}Sn, belonging to the $\nu 2f^2_{7/2}$ multiplet and one corresponding to the $\nu 2f_{7/2}1h_{9/2}$ configuration, are well reproduced by available shell-model calculations. \par
The experimental information obtained in~this work resulted in~a~significant expansion of the level scheme~of \textsuperscript{134}Sn, including seven new states, of which one is~tentatively proposed. Four levels were placed in~the range of $2-3$\,MeV, where calculations indicate the existence of members of the $\nu 2f_{7/2} 3p_{3/2}$ and $\nu 2f_{7/2}1h_{9/2}$ multiplets~\cite{10.1103/PhysRevC.84.044324, 10.1103/PhysRevC.76.024313, 10.1016/j.physletb.2016.09.030, 10.1088/1742-6596/267/1/012019}. 
The interpretation of levels at excitation energies around 5\,MeV differs for the various calculations. These differences are mainly due to the chosen model space. The calculations presented in~Ref.~\cite{10.1103/PhysRevC.84.044324} (shown in~Fig.~\ref{fig:134Sn_level_scheme_exp_vs_th_comparison}a) do~not include the $\nu 1i_{13/2}$ orbital in~the model space, but they do include core excitations by considering the $\nu 1h_{11/2}$ and $\nu 2d_{3/2}$ orbitals below the $N=82$ shell gap. 
Excited states predicted above 5~MeV belong to core-excited states with a~dominant $\nu  2f_{7/2}^3h_{11/2}^{-1}$ configuration. These are out of the model spaces of Refs.~\cite{10.1103/PhysRevC.65.051306, 10.1088/1742-6596/267/1/012019, 10.1016/j.physletb.2016.09.030, 10.1103/PhysRevC.76.024313}, which adopt a~neutron valence space consisting of orbitals above the $N=82$ shell gap only. At excitation energies exceeding 3.2\,MeV~\cite{10.1103/PhysRevC.65.051306}, 3.5\,MeV~\cite{10.1016/j.physletb.2016.09.030} and 4\,MeV~\cite{10.1103/PhysRevC.76.024313}, respectively, they predict states of negative parity that arise from particle excitations, belonging to the $\nu 2f_{7/2} 1i_{13/2}$ configuration. \par
\begin{figure}
\includegraphics[width=\linewidth]{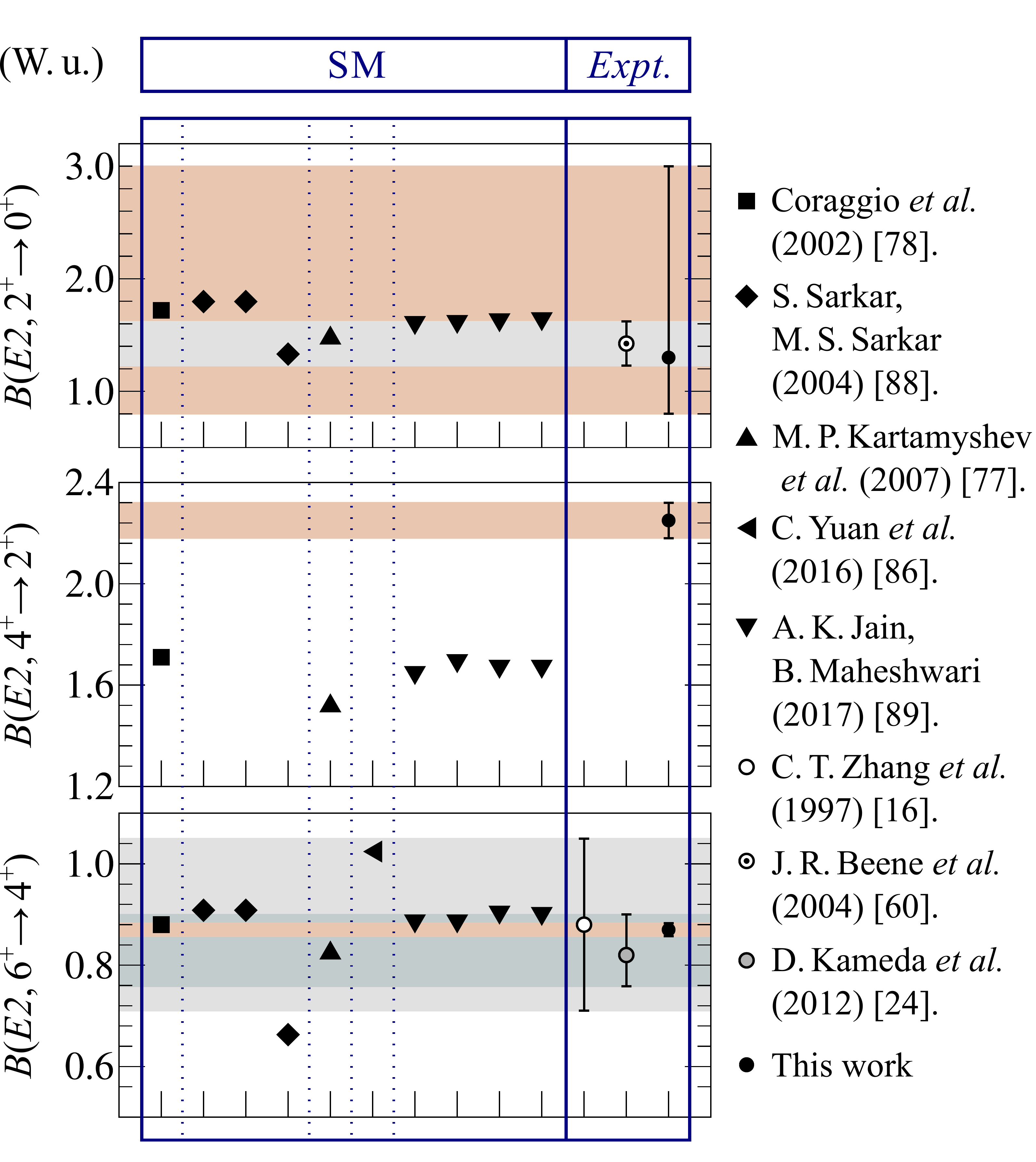}
\caption{
(Color online) 
Comparison of~predicted~(SM) and experimental~(\textit{Expt.}) reduced transition probabilities $B(E2)$ (in~W.\,u.) for~$E2$ transitions in~\textsuperscript{134}Sn. Presented data are taken from Refs.~\cite{10.1103/PhysRevC.65.051306, 10.1140/epja/i2003-10198-7, 10.1016/j.physletb.2016.09.030, 10.1088/1402-4896/aa7353, 10.1103/PhysRevC.76.024313, 10.1007/s002180050269, 10.1016/j.nuclphysa.2004.09.143, 10.1103/PhysRevC.86.054319}. 
%
Uncertainties of~the previously reported experimental results and the~one obtained in~this work are~shown by~the gray and~orange areas, respectively. 
}
\label{fig:134Sn_BE2_th_VS_exp}
\end{figure}

Reduced transition probabilities for~$E2$ transitions in~\textsuperscript{134}Sn were calculated
from the measured lifetimes of~the $2^+$, $4^+$ and $6^+$ levels. Figure~\ref{fig:134Sn_BE2_th_VS_exp} shows the~comparison of the~determined values with those reported previously and with theoretical predictions. 
The~obtained $B(E2;2^+\rightarrow\,0^+)=1.3^{+1.7}_{-0.5}$~W.\,u., is~in~agreement with the previously reported, more precise, $B(E2)$~value from~the Coulomb excitation~\cite{10.1016/j.nuclphysa.2004.09.143}, which is~well reproduced by~the shell-model calculations. The~experimental $B(E2;4^+ \rightarrow 2^+)=2.25(7)$~W.\,u., which~was measured for~the first time in~this work, is~not reproduced by~any of~the available calculations, which~consistently predict a~value of~about $1.6$~W.u., similar to~the $B(E2;2^+\rightarrow\,0^+)$ rate. In~the case of~the $6^+ \rightarrow 4^+$ transition, the~precision of~the new experimental result, $B(E2;6^+ \rightarrow 4^+)=0.870(13)$~W.\,u., is~significantly improved compared with~earlier results~\cite{10.1007/s002180050269, 10.1103/PhysRevC.86.054319}. For~this~transition rate, agreement was~obtained with various variants of~the shell-model predictions (see Figure~\ref{fig:134Sn_BE2_th_VS_exp}).
\par
If~we review the experimental and predicted trends of~$B(E2)$ values for successive transitions between members of~the $\nu 2f_{7/2}^2$ multiplet in~\textsuperscript{134}Sn, we~find that the calculations do not predict such an increase in~$B(E2)$ for the $4^+ \rightarrow 2^+$ transition as it~was observed. A~similar trend, although more pronounced, occurs for~$E2$ transitions connecting states belonging to~the analogous multiplet in~the \textsuperscript{208}Pb~region, $\nu 2g_{9/2}^2 $ in~\textsuperscript{210}Pb~\cite{10.1016/j.nds.2014.09.004, 10.1103/PhysRevC.98.024324}. 
\par
\subsubsection*{\textsuperscript{135}Sn}
Shell-model calculations for \textsuperscript{135}Sn~\cite{10.1103/PhysRevC.65.051306, 10.1140/epja/i2003-10198-7,
10.1103/PhysRevC.76.024313, 10.1103/PhysRevC.76.054305, 10.1016/j.physletb.2016.09.030} provide guidance in the
interpretation of the first experimental results on excited states for this nucleus. They predict a~$7/2^-$ spin-parity for the
ground state of \textsuperscript{135}Sn, being a~member of the $\nu 2f_{7/2}^{3}$ multiplet. This prediction is~also
supported by the systematics of excitation energies in~the $N=85$ isotones~\cite{10.1103/PhysRevC.66.044302} as well as
by the expected analogy to the \textsuperscript{133}Sn nucleus, with a~$7/2^-$ ground
state~\cite{10.1038/nature09048}. \par
\begin{figure*}
\includegraphics[width=\linewidth]{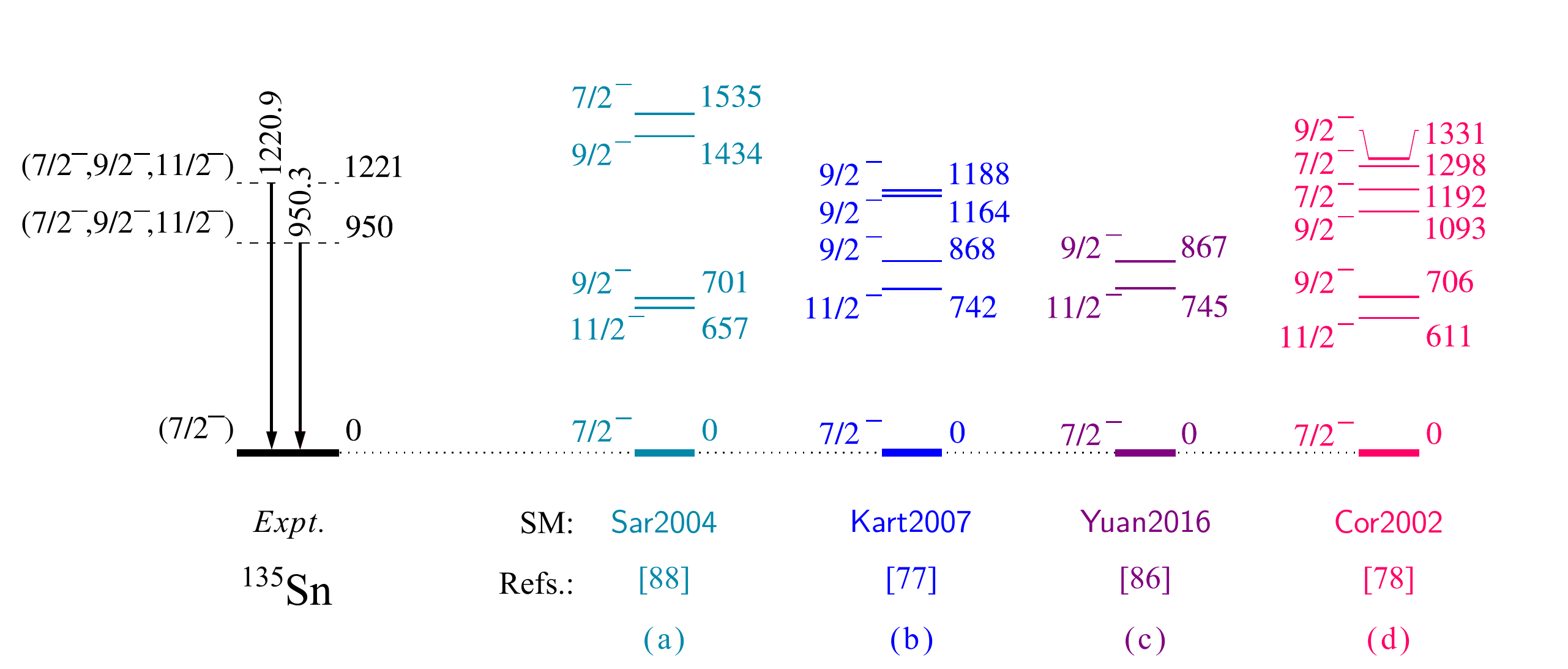}
\caption{Excited states in~\textsuperscript{135}Sn tentatively proposed in~this work (\textit{Expt.}). Calculated excitation energies (SM) for the~$7/2^-$, $9/2^-$ and $11/2^-$ states in~\textsuperscript{135}Sn reported in~Refs.: (a)~\textsf{Sar2004}~\cite{10.1140/epja/i2003-10198-7}, (b)~\textsf{Kart2007}~\cite{10.1103/PhysRevC.76.024313}, (c)~\textsf{Yuan2016}~\cite{10.1016/j.physletb.2016.09.030} and~(d)~\textsf{Cor2002}~\cite{10.1103/PhysRevC.65.051306} are~also presented. Excitation energies relative to~the \textsuperscript{135}Sn ground state are~given in~keV. The~ground-state spin-parity assignment for~\textsuperscript{135}Sn, based on~systematic trends in~neighboring nuclei, was~taken from Ref.~\cite{10.1088/1674-1137/abddae}.
}
\label{fig:135Sn_levels_th_comparison_selected}
\end{figure*}

The $5/2^-$ and~$3/2^-$~levels are~predicted to be~the lowest-lying excited states in~\textsuperscript{135}Sn~\cite{10.1103/PhysRevC.65.051306, 10.1140/epja/i2003-10198-7, 10.1103/PhysRevC.76.024313, 10.1103/PhysRevC.76.054305, 10.1016/j.physletb.2016.09.030}. Given the expected $9/2^+$~ground-state spin-parity for~\textsuperscript{135}In, their~population in~the \textsuperscript{135}In $\beta$~decay is~unlikely. 
States populated in~\textsuperscript{135}Sn via~\emph{ff}~transitions most likely have spins and~parities $7/2^-$, $9/2^-$ or~$11/2^-$. 
Figure~\ref{fig:135Sn_levels_th_comparison_selected} displays the calculated excitation energies for low-lying $7/2^-$, $9/2^-$ and $11/2^-$ levels in~\textsuperscript{135}Sn. Shell-model calculations support the tentative assignment of the 950- and 1221-keV transitions to \textsuperscript{135}Sn as~ground-state transitions, since states with such spin values are~expected in~a~comparable energy range~\cite{10.1103/PhysRevC.65.051306, 10.1140/epja/i2003-10198-7, 10.1103/PhysRevC.76.024313, 10.1103/PhysRevC.76.054305, 10.1016/j.physletb.2016.09.030}. Theoretical predictions tend to~disagree when we~consider levels at~higher excitation energies in~\textsuperscript{135}Sn, arising from the~$\nu 2f_{7/2}^{3}$, $2f_{7/2}^{2} \, 3p_{3/2}$ and~$\nu \,2f_{7/2}^{2} \, 1h_{9/2}$ configurations (see~Fig.~\ref{fig:135Sn_levels_th_comparison_selected})~\cite{10.1103/PhysRevC.65.051306, 10.1140/epja/i2003-10198-7, 10.1103/PhysRevC.76.024313, 10.1103/PhysRevC.76.054305, 10.1016/j.physletb.2016.09.030}. 
\section{Summary and conclusions}
We report on~new $\gamma$-ray spectroscopy results from the~ISOLDE facility at~CERN on~the $\beta$~decay of~the neutron-rich \textsuperscript{134}In and~\textsuperscript{135}In nuclei, populating excited states in~tin isotopes with~$N\geq 82$. Due~to~the relatively simple structure of~daughter nuclei, these $\beta$~decays provide unique conditions for~the simultaneous investigation of one- and two-neutron excitations as~well as~states formed by~couplings of~valence neutrons to~excitations of~the \textsuperscript{132}Sn~core. 
\par
The~$\beta \gamma$- and~$\beta 2n$-decay branches of~\textsuperscript{134}In have been observed for~the first time. The~$\beta$-decay scheme of~\textsuperscript{134}In was supplemented by~thirteen transitions, of~which three depopulate new~levels in~\textsuperscript{134}Sn and~two depopulate new levels in~\textsuperscript{133}Sn. Although the prevalent $\nu 1g_{7/2} \rightarrow \pi 1g_{9/2}$ GT~transition feeds neutron-unbound states at excitation energies exceeding $S_{2n}$ of \textsuperscript{134}Sn, the~\textsuperscript{134}In $\beta$~decay is~dominated by~$\beta 1n$ emission, with a~probability of~$P_{1n}=89(3)\%$. Among the available global calculations of~$\beta n$ branching ratios, only the~QRPA+HF~\cite{10.1016/j.adt.2018.03.003} and EDM~\cite{10.1103/PhysRevC.88.041301, 10.1103/PhysRevC.90.054306} models predict the predominance of~this $\beta$-decay branch for~\textsuperscript{134}In. These two theoretical approaches take into account the competition between one- and multiple-neutron emission as~well as~$\gamma$-ray deexcitation in~the decay of neutron-unbound states, which is~not included in~the other models considered. \par
A~significant contribution of~$\gamma$-ray emission from neutron-unbound states populated in~the two daughter nuclei, \textsuperscript{133}Sn and~\textsuperscript{134}Sn, at~excitation energies exceeding $S_{1n}$ by~1~MeV was observed in~this work. The~competition of~$\gamma$-ray deexcitation with neutron emission well above $S_{1n}$ can be explained by~the weak overlap of~the the wave functions of~states involved in~$\beta n$ emission. Neutron-unbound states emitting $\gamma$~rays in~\textsuperscript{134}Sn are~formed by~couplings of~valence neutrons to~core excitations, while the low-lying levels in~\textsuperscript{133}Sn arise from one-particle excitations of~valence neutron.
In~the energy range consistent with the~predicted excitation energy of~the $13/2^+$ state in~\textsuperscript{133}Sn, a~2434-keV transition was observed, which is~proposed as a~candidate for a~$\gamma$~ray depopulating the missing $\nu1i_{13/2}$ s.\,p. state  in~\textsuperscript{133}Sn.
\par
Transitions following the $\beta$~decay of~\textsuperscript{135}In were identified for~the first time and~the partial $\beta$-decay scheme of~this nucleus was established. Three~new transitions were assigned to~\textsuperscript{134}Sn based on~$\beta \gamma \gamma$ coincidences. Two~transitions were tentatively attributed to~\textsuperscript{135}Sn. Their placement in~the level scheme of~\textsuperscript{135}Sn is~supported by shell-model calculations. Several other $\gamma$~rays were observed in~the \textsuperscript{135}In $\beta$~decay but could not be assigned to a~specific $\beta$-decay branch of~the parent nucleus. Due~to their low energies and lack of~$\beta \gamma\gamma$ coincidence relations, they~cannot be placed in~the level scheme of~any other daughter nuclei. 
\par
The level scheme of~\textsuperscript{134}Sn was supplemented in~total by six new excited states, populated either through \emph{ff}~decays of~\textsuperscript{134}In or~via~$\beta 1n$ emission from~neutron-unbound states in~\textsuperscript{135}Sn. Data~from these two $\beta$~decays also allowed us~to determine the lifetimes of the previously known $2^+$,  $4^+$ and  $6^+$ states in~\textsuperscript{134}Sn. Experimental excitation energies and~reduced transition probabilities were compared with the~shell-model calculations for~\textsuperscript{134}Sn. New~levels appear~at excitation energies for~which existence of~the $\nu 2f_{7/2} 3p_{3/2}$ and~$\nu 2f_{7/2} 1h_{9/2}$~multiplets is~predicted. Calculations including core excitations reproduce well~the energies of~the two neutron-unbound states identified in~\textsuperscript{134}Sn that are~most likely populated in~the \emph{ff} $\nu 1h_{11/2} \rightarrow \pi 1g_{9/2}$ decays of~\textsuperscript{134}In.
\par \medskip
%
\section{Acknowledgements}
M.P.S.~acknowledges the~funding support from the Polish National Science Center under Grants No.~2019/33/N/ST2/03023 and No.~2020/36/T/ST2/00547 (Doctoral scholarship ETIUDA). 
J.B.~acknowledges support from the Universidad Complutense de Madrid under the Predoctoral Grant No.~CT27/16-CT28/16. 
This work was partially funded by the Polish National Science Center under Grants No.~2020/39/B/ST2/02346, 2015/18/E/ST2/00217 and 2015/18/M/ST2/00523, by the Spanish government via Projects No.~FPA2017-87568-P, RTI2018-09886~8-B-I00, PID2019-104390GB-I00 and PID2019-104714GB-C21, by~the U.K. Science and Technology Facilities Council (STFC), the German BMBF under contract 05P18PKCIA, by the Portuguese FCT under the projects~CERN/FIS-PAR/0005/2017 and CERN/FIS-TEC/0003/2019, and by the Romanian IFA Grant CERN/ISOLDE.
The research leading to these results has received funding from the European Union's Horizon 2020 research and innovation programme under grant agreement no.~654002.
M.Str. acknowledges the funding from the European Union's Horizon 2020 research and innovation program under grant agreement No.~771036 (ERC CoG MAIDEN). 
J.P. acknowledges support from the Academy of Finland (Finland) with Grant No.~307685.
Work at the University of York was supported under STFC grants ST/L005727/1 and ST/P003885/1.\par

\bibliographystyle{ieeetr}

\end{document}